%% file: main.tex
\documentclass{article}

\usepackage{arxiv}

\usepackage[utf8]{inputenc} 
\usepackage[T1]{fontenc}    
\usepackage{url}            
\usepackage{booktabs}       
\usepackage{nicefrac}       
\usepackage{microtype}      
\usepackage{lipsum}
\usepackage{graphicx}

\usepackage[numbers]{natbib}
\usepackage{mathptmx}

\usepackage{multicol}

\usepackage{amssymb,amsmath,amsfonts,amsbsy}
\usepackage{numprint}
\usepackage{tabularx}
\usepackage{setspace}
\usepackage{authblk}
\usepackage{longtable}
\usepackage{rotating}
\usepackage{float}
\usepackage{censor,caption}
\usepackage[dvipsnames]{xcolor}
\usepackage{geometry}
\usepackage{listings}
\usepackage{multirow}

\usepackage{hyperref}       
\usepackage{cleveref}
\usepackage{mycustomcommands}

\providecommand{\chg}[1]{#1}
\providecommand{\shorttitle}[1]{}
\newcommand{\savedauthornote}{}
\newcommand{\authornote}[1]{\renewcommand{\savedauthornote}{#1}}
\newcommand{\printauthornote}{%
  \begingroup
  \renewcommand{\thefootnote}{}%
  \footnotetext{\textbf{Author Note.} \savedauthornote}%
  \endgroup
}\graphicspath{ {./images/} }
\usepackage{tikz}
\tikzstyle{ov}=[shape=rectangle,draw=black!80,minimum height=0.6cm,
                minimum width=0.6cm,thick]
\tikzstyle{lv}=[shape=circle,draw=black!80,thick,minimum width=0.7cm]
\tikzstyle{lr}=[shape=circle,draw=black!00,thick,minimum width=0.7cm]
\tikzstyle{re}=[shape=circle,draw=black!80,thick,minimum width=0.7cm,dotted]

\title{Dynamic Latent Class Structural Equation Modeling: A Hands-On Tutorial for Modeling Intensive Longitudinal Data}
\shorttitle{DLCSEM Tutorial}

\author[1]{Roberto Faleh\textsuperscript{*}}
\author[1]{Sofia Morelli}
\author[1]{Vivato Andriamiarana}
\author[2]{Zachary J. Roman}
\author[3]{Christoph Fl\"uckiger}
\author[1]{Holger Brandt}

\affil[1]{University of T\"ubingen, Methods Center}
\affil[2]{University of Z\"urich, Department of Informatics}
\affil[3]{University of Kassel, Department of Psychology}

\authornote{
Roberto Faleh and Sofia Morelli contributed equally and share first authorship.
Data and reproducible code are available at \url{https://github.com/PsychometricsMZ/dsem_tutorial}.
The data were collected as part of the IMPLEMENT randomized controlled trial \cite{Flueckiger2016} and have been previously described in \citet{Flueckiger2021}.\newline
\textsuperscript{*}Corresponding author: \texttt{roberto.faleh[at]uni-tuebingen.de}.
}

\begin{document}

\maketitle
\printauthornote

\begin{abstract}

In this tutorial, we provide a hands-on guideline on how to implement complex Dynamic Latent Class Structural Equation Models (DLCSEM) in the Bayesian software JAGS. We provide building blocks starting with simple Confirmatory Factor and Time Series analysis, and then extend these blocks to Multilevel Models and Dynamic Structural Equation Models (DSEM). \chg{Subsequently, we introduce Hidden Markov Switching Models (HMSM) and demonstrate their integration with DSEM to yield DLCSEM.} Leading through the tutorial is an example from clinical psychology using data on a generalized anxiety treatment that includes scales on anxiety symptoms and the Working Alliance Inventory that measures alliance between therapists and patients. Within each block, we provide an overview, specific hypotheses we want to test, the resulting model and its implementation, as well as an interpretation of the results. The aim of this tutorial is to provide a step-by-step guide for applied researchers that enables them to use this flexible DLCSEM framework for their own analyses.
\end{abstract}

\newpage

\newpage
\input{00a_intro}
\input{00b_strategy}
\input{01_cfa}
\input{02_timeseries}
\input{03_mlm}
\input{04_dsem}
\input{05_hmsm}
\input{06_dlcsem}
\input{07_fusion}
\input{08_discussion}
\bibliography{literature}

\appendix

\input{app02_notation}

\input{app03_convergence}
\input{app04_fusion}

\end{document}

%% file: 00a_intro.tex
\section{Introduction}
Recent advances in Bayesian latent variable modeling, namely Dynamic Structural Equation Modeling \citep[DSEM;][]{Asparouhov2018} and its extension to Dynamic Latent Class Structural Equation Modeling \citep[DLCSEM;][]{Asparouhov2017, Kelava2019}, have provided powerful tools for analyzing intensive longitudinal data (ILD). 

ILD capture information from multiple subjects observed over numerous time intervals, ranging from thirty to several hundred distinct time points \cite{Bolger2013}. Often, ILD include information from several sources, such as self-reports, physiological, behavioral, environmental, and geographic data \citep[e.g.,][]{Ginexi2014}. Such intensive real-time data collection in natural environments is commonly referred to as ambulatory assessment \cite{Trull2014}. Within intervention research, this approach is frequently implemented as Ecological Momentary Assessment \citep[EMA;][]{Heron2010}.

The growing availability of mobile phones, smartwatches, and other personal wearable devices has made ILD collection more accessible and cost-effective \cite{Riley2015}. Compared to traditional intervention studies with only a few measurement occasions, the collection of ILD via EMA offers several advantages. Because data are obtained in real time and natural settings, self-reports are less affected by retrospective recall biases, thereby increasing generalizability and ecological validity \cite{Stone2007}. Given these advantages, ILD have become an important data source for researchers in clinical and health psychology \citep[e.g.,][]{Jonasson2019, Scholz2019}.

The richness of ILD enables researchers to capture both intra- and inter-individual processes and dynamic changes, enhancing the ability to predict future events and improve individual forecasts. This is particularly important for identifying potential relapse triggers or the conditions necessary for optimal treatment efficacy \citep[e.g.,][]{Businelle2016}. DSEM and DLCSEM effectively model such complex processes in ILD, enabling researchers to capture intra-individual change, heterogeneity in treatment response, and latent state switching over time.

DSEM can be viewed as a combination of time-series models and multilevel SEM. The time-series model on the within-level can be set up for both observed and latent variables. It can include all types of auto-regressive (AR) and moving-average (MA) specifications. 
Each within-level coefficient can be specified as a random effect to account for person- and time-specific effects. Person-specific random effects provide information about individual differences. This information can produce more precise forecasts of individual trajectories and, for example, improve the design of tailored interventions. Time-specific random effects describe dynamical (e.g., accelerated) changes over time \cite{Fruehwirth2008}. The between model can include further information, such as (observed or latent) baseline covariates.

DLCSEM builds upon the foundation established by DSEM but introduces a key extension:  a latent class model that captures categorical shifts over time, referred to as 'states'. In traditional Latent Class Analysis (LCA) and its longitudinal extension through Growth Mixture Modeling (GMM), including Latent Class Growth Analysis (LCGA), class membership is assumed to remain static over time \cite{Muthen2000, Jung2008}. DLCSEM, in contrast, allows state membership to change dynamically, thereby enabling individuals to follow different state-specific processes at each time point. Unlike original Latent Transition Analysis \citep[LTA;][]{Collins1992, Lanza2003}, which assumes homogeneous transition probabilities across individuals, DLCSEM further permits individual-specific transition probabilities between latent classes over time by incorporating random effects. A comparison of these approaches is provided in \Cref{tab:model-comparison}.

\input{app01_comparison_table}

The ability of DLCSEM to identify critical states, detect stable a priori differences, account for unobserved heterogeneity, and capture intra-individual changes extends the capabilities of traditional dynamic modeling. It enables real-time dynamic classification of individual states, providing a powerful tool for personalized evaluations and prompt interventions. Moreover, it offers aggregated results comparable to those obtained from DSEM, supporting the evaluation of treatments while also classifying individuals into distinct states based on their dynamic responses.These modeling flexibilities provide a detailed representation of temporal dynamics and deep insight into the underlying dynamical process.

\subsection{Applications of DLCSEM in the Literature}

\citet{Flueckiger2021} applied DLCSEM to investigate changes in the dimensional structure of the Working Alliance Inventory \citep[WAI;][]{Horvath1989} over time. They found that patients' response patterns shifted during the course of a psychotherapeutic intervention. While patients initially responded according to a three-dimensional scale representing the subfacets of the working alliance, most began to follow a single-dimensional scale (global factor) in later sessions. This phenomenon, referred to as the fusion of subscales, can be analyzed using DLCSEM.

In \citet{Kelava2022}, the framework was used to investigate student dropout from university. Their findings revealed that the latent state variable acted as a proxy for students' intention to drop out. This variable was shown to predict actual dropout, which typically occurred, on average, six measurement occasions after the estimated intention to drop out. Additionally, they identified relevant time-varying covariates, such as negative affect, which significantly predicted this shift beyond baseline covariates like cognitive skills.

\citet{Roman2024} used DLCSEM to identify inattentive response patterns in long questionnaires. In this application, one latent state represented the response pattern for attentive participants using a hypothesized factor structure, and the second state modeled inattentive response behavior using person-specific parameters that are independent of the underlying constructs that were supposed to be measured.

What is common in all of these examples is that (a) a psychometric theory is used to define the latent variables, and (b) a confirmatory approach is taken to specify the different states. This approach is different from the often exploratory use of latent class models in the past. While some researchers have already argued for a confirmatory application of latent class models \cite{Jeon2019}, it is even more relevant for DLCSEM. This is due to two reasons: First, the complexity of the DLCSEM framework is such that an exploratory search for latent states is unfeasible because each parameter can be state-specific. Second, convergence and parameter bias heavily depend on a thorough state specification.

\subsection{Scope and Outline}

In this tutorial, we provide a step-by-step guide to implementing and interpreting Bayesian latent variable models for ILD, starting with foundational concepts and gradually building toward more complex frameworks. The aim is to offer a practical toolbox that equips readers with the concepts and techniques needed to understand, apply, and interpret DLCSEM. While the tutorial is largely self-contained, readers are assumed to have some familiarity with factor analysis \citep[for an introduction, see e.g.,][]{Bollen2002,Loehlin2003}, time series analysis \citep[e.g.,][]{Box2015,Shumway2025}, and basic concepts of Bayesian inference, including prior specification, convergence assessment, and posterior interpretation \citep[e.g.,][]{McElreath2020,Gelman2013}, as well as sufficient experience with \textsf{R} \cite{Rlanguage} to read \textsf{JAGS} code \cite{Plummer2003}.

The sections are organized as follows: We first discuss the basic idea of this stepwise tutorial and introduce data, hypotheses, and implementation background. Then, we move on with Bayesian implementations of CFA, Time Series Models, and Multilevel Models, before combining them into a DSEM. We then explain HMSMs and finally show how they can be combined with DSEM to form DLCSEM. Throughout, we follow a didactic structure, beginning with familiar, simpler models (e.g., factor models) and introducing key building blocks that are then extended to DLCSEM. This approach allows researchers to adapt more complex models based on their prior experience.

The stepwise procedure supports researchers by developing new and interesting hypotheses; they learn how to translate these hypotheses into DLCSEM and how to interpret the output with regard to their hypotheses. They can make themselves familiar with several steps they should take in the analysis strategy that will support a valid interpretation of the results.

%% file: app01_comparison_table.tex
\begin{table}[H]

\caption{Conceptual comparison of DSEM, latent class models, and DLCSEM.}
\centering
\begin{tabular}{lccccc}
\toprule
 & DSEM & LCA & LCGA/GMM & LTA & DLCSEM \\
\midrule
Latent variables (SEM framework)          & \checkmark &            &            &            & \checkmark \\
Latent classes                             &            & \checkmark & \checkmark & \checkmark & \checkmark \\
Longitudinal dynamics (change over time)                             & \checkmark &            & \checkmark & \checkmark & \checkmark \\
Class membership fixed over time           &            & \checkmark & \checkmark &            &            \\
Class membership can change (switching)    &            &            &            & \checkmark & \checkmark \\
Random effects (multilevel structure)      & \checkmark &            & \checkmark &            & \checkmark \\
Individual-specific transition dynamics    &            &            &            &            & \checkmark \\
\bottomrule
\end{tabular}

\vspace{0.5em}
\begin{minipage}{0.95\textwidth}
\footnotesize\textit{Notes.} DSEM: \cite{Asparouhov2018}. LCA: \cite{Muthen2000}. \newline
LCGA/GMM: \cite{Jung2008}. LTA: \cite{Collins1992,Lanza2003}.\newline
DLCSEM: \cite{Kelava2019}.
\end{minipage}

\label{tab:model-comparison}
\end{table}

%% file: 00b_strategy.tex
\newpage
\section{Modeling Strategy and Background Information for the Tutorial}

\chg{The modular structure of DLCSEM presented in this tutorial builds on practical guidelines for DSEM \cite{Asparouhov2022, Hamaker2021} and extends them with recommendations tailored to DLCSEM. Specifically, we propose the following stepwise modeling strategy:}

\chg{\begin{enumerate}
    \item \textbf{Begin with CFA at separate time points.} Assess the fit of the basic measurement model before moving to dynamic analyses. A poorly measured latent construct can propagate uncertainty into subsequent model layers, potentially resulting in convergence problems and unstable estimates.
    \item \textbf{Explore DSEM time‐series and MLM specifications incrementally.} Evaluate different lag structures and random‐effects configurations to determine an appropriate level of model complexity before introducing HMSM.
    \item \textbf{Incorporate the HMSM component to obtain the DLCSEM.} This step is particularly delicate and requires carefully formulated hypotheses that are properly integrated into the model. Ill-defined switching mechanisms can easily lead to convergence difficulties, unreliable estimates, or invalid interpretations.
\end{enumerate}}

\chg{These guidelines provide a road map for building a valid DLCSEM analysis. In the remainder of the tutorial, we discuss each modeling step using illustrative real‐data examples and highlight key takeaways that link the examples to the general modeling procedure. We provide fully commented code for all examples in our GitHub repository:\newline
\url{https://github.com/PsychometricsMZ/dsem_tutorial}.}

\subsection{Data}

The data set used in this tutorial is drawn from a randomized controlled implementation trial for cognitive behavioral therapy for generalized anxiety \citep[IMPLEMENT;][]{Flueckiger2016}. A detailed description can be found in the original article, and in \citet{Flueckiger2021}.\footnote{For the study protocols, refer to \citet{Flueckiger2014protocol, Flueckiger2018protocol}. Both trials were preregistered at ClinicalTrials.gov (NCT02039193 and NCT03079336) and approved by the respective Ethics Committees (see KEK 2011-0475 and BASEC 2016-00773).}

For the illustrations in this tutorial, we included all $57$ patients and $15$ time points. The majority of the patients were female (75\%), and about half of the patients showed comorbidity with additional psychological disorders. 
Treatment was conducted in a 50–60 min session format that consisted of 14 sessions and one booster session, and was based on the MAW-workbook \cite{Craske2006} that each patient received.

The main outcomes considered here were based on the Beck Anxiety Inventory \citep[BAI;][]{Beck1988} and the Working Alliance Inventory \citep[WAI;][]{Horvath1989, Munder2010}; all scales were assessed at each session. 
The BAI consisted of 21 items of general anxiety symptoms, which the patient completed before each session. It was used as a symptom measure in the IMPLEMENT trial. For the analyses here, we comprised the 21 items into 3 parcels \citep[again, see][]{Flueckiger2021}. 
The WAI was developed to measure each of the three focal elements of working alliance (task, goals, and bonds). It was used in its German translation of the WAI—Revised Short Form \cite{Munder2010}, which consists of 12 items with a 5-point Likert. Here, we used 3 per subscale in line with \citet{Flueckiger2021}. \chg{Each variable was rescaled prior to the analyses using mean and standard deviation of the first measurement occasion. This scaling keeps relative differences intact, but moves the scale to standardized variables at time point 1.}

\subsection{Hypotheses}

\chg{In this tutorial, we aim to explore and test the following hypotheses, which guide our illustrative investigations:}

\begin{enumerate}
    \item[H1] \chg{The measurement models specified for the considered scales demonstrate configural measurement invariance across time, with the same latent factor structures providing an adequate representation of the data at both the initial and final measurement occasions.}
    \item[H2] \chg{Anxiety development exhibits interindividual differences.}
    \item[H3] \chg{Anxiety evolves dynamically, with temporal dependencies that vary across measurement occasions.}
    \item[H4] \chg{Anxiety levels exhibit abrupt within-person changes over time.}
    \item[H5] \chg{The factor structure of the WAI scale shows a dimensional shift.}
\end{enumerate}

\chg{We will informally test H1 using cross-sectional CFA to examine factor structures at selected time points. H2 will be assessed using MLM and DSEM by incorporating person-specific random intercepts and slopes to capture interindividual differences. H3 will be addressed using DSEM with a time-varying autoregressive effect ($\operatorname{AR}(1)$ structure with a time-specific random slope) to model temporal variation in dependencies. H4 and H5 will be formulated within the DLCSEM framework: the model for H4 specifies an HMSM in which states represent different levels of anxiety but share a similar underlying structure, whereas the model for H5 allows structurally distinct models in each state.}

\subsection{Analysis}

All analyses were conducted in R \chg{\cite{Rlanguage}} with models specified in JAGS \cite{Plummer2003}. \chg{JAGS is a well-maintained, open-source software with a concise and expressive syntax that integrates seamlessly with R and runs reliably across platforms, providing a free and accessible alternative to licensed software such as Mplus \cite{Muthen1998mplus}.}

\chg{We rely on a Bayesian approach primarily because it allows convenient specification of models through their full set of conditional distributions \citep[see Appendix~B of][for details in the DSEM context]{Asparouhov2018}, which have long been established in the literature \cite{Arminger1998} and can be directly implemented in a Gibbs sampler \cite{Gelfand1990}, the core sampling mechanism of the Monte Carlo Markov Chain (MCMC) algorithm used by JAGS. By contrast, a frequentist implementation would require constructing a high-dimensional joint likelihood, which is analytically challenging and often impractical for novel model extensions \cite{Asparouhov2018}.}

\chg{The Bayesian framework also allows prior knowledge to be incorporated transparently and coherently. Priors enable researchers to integrate existing evidence, encode theoretical expectations, and express uncertainty. Throughout this tutorial, we use weakly informative priors in line with \citet{Kelava2019} unless otherwise specified. Priors for newly introduced parameters are described in the corresponding section and retained for the remaining models unless noted. A detailed discussion of prior choices, especially with regard to sample size and convergence, is provided in Appendix~\ref{app:convergence}.}

\chg{We ran each model with four chains, drawing a total of 20{,}000 posterior samples (10{,}000 iterations per chain, with the first 5{,}000 discarded as burn-in) prioritising computational efficiency over precision. Convergence was assessed using the $\hat{R}$ statistic \cite{GelmanRubin1992, BrooksGelman1998}, effective sample sizes, and visual inspection of trace and density plots.}

\chg{Concerning model structure, we assume equal time intervals and impose homoskedasticity and measurement invariance, unless stated otherwise. These assumptions are revisited in the final discussion.}

%% file: 01_cfa.tex
\section{Confirmatory Factor Analysis (CFA)}

\chg{Before turning to ILD-specific time-series analyses, we first demonstrate how CFA can be used to assess factor dimensionality at selected time points. As discussed at the end of this section, this step provides a general strategy to inform the specification of a full dynamic model. To illustrate this approach, we apply Bayesian CFA \cite{Lee1981} to cross-sectional slices of the data and informally examine configural measurement invariance.}

\subsection{Data and Goal of Illustration 1}

\chg{We expect a unidimensional CFA model to adequately represent responses to the BAI items, and a three-factor CFA model with correlated task, goals, and bond factors to adequately represent responses to the WAI items. H1 posits that the BAI and WAI consistently capture their intended constructs over time. We evaluate this hypothesis heuristically by examining the factor structure at the beginning and end of the study period, comparing CFA results from the initial ($t = 1$) and final ($t = 15$) measurement occasions.}

\subsection{CFA model implementation}

The outcome $y_{ij}$ of each patient  $i\in(1,...,N)$ for indicator $j\in(1,...,12)$ is assumed to follow a normal distribution $\mathcal{N}$ with mean $\mu_{y_{ij}}$ and item-specific residual variance $\sigma_{y_j}^2$.
In the JAGS code, the second parameter must be specified as precision \texttt{psi.y[j]} ($=\sigma_{y_j}^{-2}$) due to the implementation of the normal distribution \texttt{dnorm}:
\numberedtabx{
$y_{ij}\sim \mathcal{N}(\mu_{y_{ij}},\sigma_{y_j}^2)$  
    &\quad\quad\texttt{y[i,j] $\sim$ dnorm(mu.y[i,j], psi.y[j])}.
}{code:cfa4factorOutcome}
\chg{Assuming a simple structure and adding standard scaling constraints for identification, the mean structure for the first factor is given by}:
\numberedtabx{
$\mu_{y_{i1}}=\eta_{i1}$
  &\texttt{mu.y[i,1] <-             eta[i,1]}\\
$\mu_{y_{i2}}=\nu_1+\lambda_{1}\cdot\eta_{i1}$
  &\texttt{mu.y[i,2] <- nu.y[1]+lambda.y[1]*eta[i,1]}\\
$\mu_{y_{i3}}=\nu_2+\lambda_{2}\cdot\eta_{i1}$
  &\texttt{mu.y[i,3] <- nu.y[2]+lambda.y[2]*eta[i,1]}\\
}{code:cfa4factorMeasurementModel}
\chg{where $\nu$ and $\lambda$ represent the indicators' intercept and loading}, and $\eta_{ik}$ denotes the value of the $k$-th latent factor for patient $i$.\footnote{Note that we use different indices compared to \chg{typical notation (such as $\lambda_{21}$ for indicator 2 loading on factor 1)} for the intercepts $\nu$ and factor loadings $\lambda$ to match the code. Instead of using two indices for the indicator $j$ and the factor number $k$, we count the free parameters from $1$ to $8$.}

The latent factor \chg{score vector} $\boldsymbol\eta_i$ for each patient $i$ are assumed to follow a joint multivariate normal distribution with mean vector $\boldsymbol{\mu_{\eta}}$ and covariance matrix $\boldsymbol{\Sigma}_{\eta}$\footnote{Again, in JAGS, the second parameter is specified as precision matrix \texttt{psi.eta} ($=\boldsymbol{\Sigma}_{\eta}^{-1}$).}:
\numberedtabx{
$\boldsymbol\eta_{i}\sim \mathcal{MVN}(\boldsymbol{\mu_{\eta}},\boldsymbol{\Sigma}_{\eta})$
  &\texttt{eta[i,1:4] $\sim$ dmnorm(mu.eta[1:4],}\\
  &\hspace{4.4cm}\texttt{psi.eta[1:4,1:4]).}\\
}{code:4factorDistribution}
The path diagram of this model is represented in \Cref{fig:01_cfa}. The priors are specified in \Cref{tab:priorsCFA}. 

\input{fig/01_cfa}

\begin{table}[H]
    \centering
    \caption{Priors for the CFA model with $4$ factors $\eta_m$ loading on $3$ indicators $y_j$ each. (In the code, the second parameter of the normal distribution is defined as the precision=1/variance.)}
    \label{tab:priorsCFA}
    \begin{tabularx}{\textwidth}{lllX}
        \hline
        Parameter & Prior family & Prior & Code\\
        \hline
        Residual & Gamma distribution & $\sigma_{y_j}^{-2} \sim \Gamma(1, 1)$ & \texttt{for (j in 1:12)\{}\\
        precision & (uninformative) &  &  \quad\texttt{psi.y[j]}\\
        & & &  \quad \quad\texttt{$\sim$ dgamma(1,1)}\\
        & & & \texttt{\}}\\
        Free factor & Censored normal  & $\lambda_j \sim \mathcal{N}^{+}(0.5,1)\,$ & \texttt{for (j in 1:8)\{}\\
        loading & distribution &  &  \quad\texttt{lambda.y[j]}\\
        & (weakly informative) & &  \quad \quad\texttt{$\sim$ dnorm(0.5,1)I(0,)}\\
        & & & \texttt{\}}\\
        Free & Normal distribution & $\nu_j \sim \mathcal{N}(0, 100)$ & \texttt{for (j in 1:8)\{}\\
        intercept & (uninformative) &  &  \quad\texttt{nu.y[j]}\\
        & & &  \quad \quad\texttt{ $\sim$ dnorm(0,0.01)}\\
        & & & \texttt{\}}\\
        Factor  & Normal distribution & $\mu_{\eta_m} \sim \mathcal{N}(0, 100)$ & \texttt{for (m in 1:4)\{}\\
         mean & (uninformative) &  & \quad\texttt{mu.eta[m]}\\
         & & & \quad\quad\texttt{$\sim$ dnorm(0,0.01)}\\
         & & & \texttt{\}}\\
        Factor & Wishart distribution & $\boldsymbol{\Sigma}_{\eta}^{-1} \sim Wish(\Sigma_{0}^{-1}, 4)$ & \texttt{psi.eta[1:4,1:4]}\\
        precision & (uninformative) & with hyperprior $\Sigma_{0}^{-1}$ & \quad\texttt{$\sim$ dwish(psi0,4)}\\
        & & & \texttt{psi0 = diag(4)}\\
        \hline\hline
    \end{tabularx}
\end{table}

\subsection{Results of Illustration 1}

\chg{At the initial time point ($t=1$), standardized factor loadings were moderate to high, ranging from 0.55 to 0.86 (first column in \Cref{tab:posterior_fact}). Factors 2 (task), 3 (goal), and 4 (bond) were positively correlated (.643 between 2 and 3, .598 between 2 and 4, and .488 between 3 and 4), while factor 1 (anxiety) had small negative correlations with the other factors (-.039, -.154, and -.209).}

\begin{table}[H]
    \centering
    \caption{Illustration 1: Posterior estimates of free factor loadings $\lambda$ from the CFA model at $t=1$ and $t=15$, including posterior mean, standard deviation (SD), and 95\% credible intervals (2.5\% and 97.5\% quantiles).}
    \label{tab:posterior_fact}
    \begin{tabular}{lcccc|cccc}
        \hline
        \multirow{2}{*}{Parameter} & \multicolumn{4}{c|}{$t=1$} & \multicolumn{4}{c}{$t=15$} \\
         & Mean & SD & 2.5\% & 97.5\% & Mean & SD & 2.5\% & 97.5\% \\
        \hline
        $\lambda_1$ & 1.02 & 0.28 & 0.53 & 1.64
        &  0.81 & 0.23 & 0.40 & 1.30 \\
        $\lambda_2$ & 1.24 & 0.27 & 0.78 & 1.85
        & 1.29 & 0.35 & 0.66 & 1.98 \\
        $\lambda_3$ & 1.34 & 0.25 & 0.92 & 1.88 
        & 1.10 & 0.12  & 0.87 & 1.36 \\
        $\lambda_4$ & 1.35 & 0.27 & 0.89 & 1.94
        & 0.79 & 0.13 & 0.54 & 1.07 \\
        $\lambda_5$ & 1.29 & 0.32 & 0.74 & 1.98
        & 0.92 & 0.17 & 0.61 & 1.31 \\
        $\lambda_6$ & 0.95 & 0.32 & 0.39 & 1.63
        & 0.86 & 0.18 & 0.53 & 1.24 \\
        $\lambda_7$ & 1.29 & 0.27 & 0.81 & 1.87
        & 0.59 & 0.12 & 0.37 & 0.85 \\
        $\lambda_8$ & 1.03 & 0.34 & 0.42 & 1.77
        & 0.71 & 0.16 & 0.42 & 1.04 \\
        \hline
    \end{tabular}
\end{table}

\chg{At the later time point ($t=15$), standardized factor loadings were slightly higher than at $t=1$, ranging from 0.64 to 0.91 (second column in \Cref{tab:posterior_fact}). Factors 2, 3, and 4 showed strong positive correlations (.823 between 2 and 3, .829 between 2 and 4, and .776 between 3 and 4), while factor 1 had weak negative correlations with the other factors (-.195, -.223, and -.112).}

\subsection{Discussion of Illustration 1}
  
\chg{With regard to hypothesis H1, the measurement models specified for the considered scales generally support configural measurement invariance across time. Factor loadings remain consistently high, providing evidence that the expected latent structures are adequately captured at both the initial and final measurement occasions. At the same time, the correlations between factors change notably: compared to the correlations at $t = 1$, those among factors 2, 3, and 4 (the WAI subfacets) at $t = 15$ are considerably higher. This pattern suggests a potential evolution in the factor structure, indicating that the three-factor WAI model may collapse into a more unified single-factor structure over time. We will further investigate this fusion of subscales with an DLCSEM in Illustration 6.}

\subsection{\chg{Learnings Regarding the Procedure}}

\chg{This initial psychometric testing illustrates the first step in the modeling guidelines: evaluating the factor structure. While examining additional time points could further refine these insights, doing so for many time points (e.g., 50) quickly becomes impractical. In such cases, selecting a subset of time points (e.g., 3 or 4) can provide approximate information, which should then be complemented by subsequent modeling steps.}

\chg{Given the largely heuristic nature of this testing procedure, it is particularly important to use scales that have been psychometrically validated rather than ad hoc implementations. If the underlying factor model is unclear, there is currently no reliable method for modeling latent variables within the context of DSEM (e.g., via exploratory factor analysis). Large shifts in the measurement model across time points can change the meaning of the latent factor and may render a time series model inappropriate.}

%% file: fig/01_cfa.tex
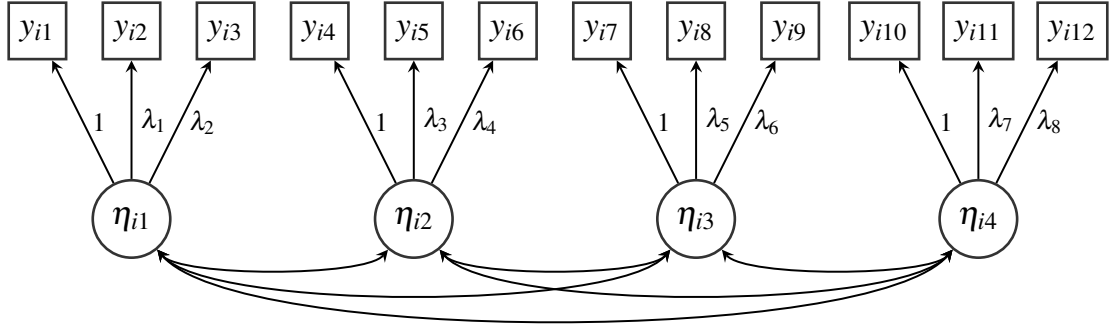
\begin{figure}[H]
\begin{center}
\resizebox{.9\textwidth}{!}{
\begin{tikzpicture}[>=stealth,semithick]

\node[ov] (y1) at (0,0)  {$y_{i1}$};
\node[ov] (y2) [right of=y1]  {$y_{i2}$};
\node[ov] (y3) [right of=y2]  {$y_{i3}$};
\node[lv, below of=y2, node distance=2cm] (eta1) {$\eta_{i1}$};

\node[ov] (y4) at (3,0)  {$y_{i4}$};
\node[ov] (y5) [right of=y4]  {$y_{i5}$};
\node[ov] (y6) [right of=y5]  {$y_{i6}$};
\node[lv, node distance=2cm] (eta2) [below of=y5]  {$\eta_{i2}$};

\node[ov] (y7) at (6,0)  {$y_{i7}$};
\node[ov] (y8) [right of=y7]  {$y_{i8}$};
\node[ov] (y9) [right of=y8]  {$y_{i9}$};
\node[lv, node distance=2cm] (eta3) [below of=y8]  {$\eta_{i3}$};

\node[ov] (y10) at (9,0)  {$y_{i10}$};
\node[ov] (y11) [right of=y10]  {$y_{i11}$};
\node[ov] (y12) [right of=y11]  {$y_{i12}$};
\node[lv, node distance=2cm] (eta4) [below of=y11]  {$\eta_{i4}$};

\draw[->] (eta1) -- node[right, scale=0.8] {1} (y1);
\draw[->] (eta1) -- node[right, scale=0.8] {$\lambda_1$} (y2);
\draw[->] (eta1) -- node[right, scale=0.8] {$\lambda_2$} (y3);
\draw[->] (eta2) -- node[right, scale=0.8] {1} (y4);
\draw[->] (eta2) -- node[right, scale=0.8] {$\lambda_3$} (y5);
\draw[->] (eta2) -- node[right, scale=0.8] {$\lambda_4$} (y6);
\draw[->] (eta3) -- node[right, scale=0.8] {1} (y7);
\draw[->] (eta3) -- node[right, scale=0.8] {$\lambda_5$} (y8);
\draw[->] (eta3) -- node[right, scale=0.8] {$\lambda_6$} (y9);
\draw[->] (eta4) -- node[right, scale=0.8] {1} (y10);
\draw[->] (eta4) -- node[right, scale=0.8] {$\lambda_7$} (y11);
\draw[->] (eta4) -- node[right, scale=0.8] {$\lambda_8$} (y12);

\path[<->] 
    (eta1)  edge [bend right=50,looseness=.4] (eta2)
    (eta1)  edge [bend right=50,looseness=.4] (eta3)
    (eta1)  edge [bend right=50,looseness=.4] (eta4)
    (eta2)  edge [bend right=50,looseness=.4] (eta3)
    (eta2)  edge [bend right=50,looseness=.4] (eta4)
    (eta3)  edge [bend right=50,looseness=.4] (eta4)
  ;
\end{tikzpicture}
}
\end{center}
\caption{Path diagram of the confirmatory factor analysis (CFA) model at a single time point for $N$ individuals. The model includes four latent factors ($\eta$), each loading onto three observed indicators ($y$) with freely estimated factor loadings ($\lambda$). Residual variances are not depicted for simplicity.}
\label{fig:01_cfa}
\end{figure}

%% file: 02_timeseries.tex
\section{Time Series Models}

\chg{Longitudinal data call for time-series models that describe how variables influence one another across successive time points, allowing us to study dynamic processes within individuals \citep[e.g.,][]{Hamaker2021}. One such approach, which we adopt in this section, is the autoregressive ($\operatorname{AR}$) model \cite{Box1968}. This time series structure is particularly suitable for behaviors driven by regular, stable internal mechanisms, including addictive behaviors or the progression of depressive symptoms \cite{Bulteel2018}.}\footnote{\chg{An alternative are moving average ($\operatorname{MA}$) models, which are more dynamic in the sense that external, occasional events affect the change process, such as in the case of occasional smokers or drinkers \citep[e.g.,][]{Velicer1996}. It is also possible to use their combination, \citep[$\operatorname{ARMA}$;][]{Box2015}, or more advanced variants such as autoregressive conditional heteroscedasticity ($\operatorname{ARCH}$) models \cite{Engle1982} and their generalized form \citep[$\operatorname{GARCH}$;][]{Bollerslev1986}. Other alternative approaches, including continuous-time models \cite{Voelkle2013} and seasonal extensions of ARMA (e.g., SARIMA; LIT), have not yet been applied to DLCSEM.}} 

A key feature of AR models is their ability to preserve a ``long memory'' by explicitly modeling dependencies across time points.
A general $\operatorname{AR}(p)$ process linearly depends on the outcomes from the previous $p$ time steps:
\begin{equation}
  \begin{aligned}
    y_t &=  \alpha+\beta_1 y_{t-1}+\ldots+\beta_p y_{t-p}+\varepsilon_t 
    \end{aligned}  
\end{equation}
where $p$ is called the order of the process. The difference between time points $t$ and $t-t_k$ is termed lag. Higher-order processes (larger $p$) include more lags, allowing the model to account for broader, direct influences of past outcomes on future values.

A first-order autoregressive process, $\operatorname{AR}(1)$, is often sufficient for modeling simple psychological phenomena \cite{Bulteel2018}. It is defined as 
\begin{equation}
    y_t= \alpha + \beta y_{t-1}+\varepsilon_t,
    \label{eq:ar1}
\end{equation}
including only one lag. The parameter $\alpha$ acts as an intercept. The parameter $\beta$ is the autoregressive coefficient determining the persistence of the memory over time.

The error term $\varepsilon_t$ introduces random ``shocks'' that propagate through time via the autoregressive structure. Even with a single lag, each step continues to propagate the influence of past errors, creating an infinite memory. If $\mid \beta \mid > 1$, then any shocks described by the white noise process $\varepsilon_t$ will accumulate over time. However, such non-stationary behavior is rare in natural systems, as feedback mechanisms often prevent unbounded growth. 
For this reason, we typically assume stationarity of the time series, with $\mid \beta \mid \ < 1$ such that the effects of past events fade over time. This can be interpreted as a gradually diminishing memory.

\subsection{Data and Goal of Illustration 2}

In the following example, we use data collected from a single patient  ($N=1$) over $N_t=15$ time steps.\footnote{Naturally, the results depend highly on the selected patient. We chose patient $8$ for demonstration because their data is complete and they experienced a visible change in anxiety level.} The goal is to introduce the implementation of an $\operatorname{AR}(1)$ time structure to model the temporal dependencies in the data, \chg{using item $1$ from the BAI as the observed outcome}.

\subsection{$\operatorname{AR}$(1) model implementation}

\chg{For the implementation of an $\operatorname{AR}$(1) process, we slightly adapt the definition in \Cref{eq:ar1}} by centering the past outcome $y_{t-1}$ around the intercept $\alpha$:
\begin{equation}
  y_t =\alpha + \beta \cdot (y_{t-1}-\alpha)+\varepsilon_{t}.
  \label{eq:centered_ar1}
\end{equation}
This adjustment eliminates the dynamic panel bias, also known as Nickell’s bias \chg{\cite{Nickell1981, Raudenbush2002}. The error term of an $\operatorname{AR}$ process is typically assumed to follow a white-noise sequence with zero mean ($E(\varepsilon_t) = 0$), constant variance ($E(\varepsilon_t^2) = \sigma^2$), and no autocorrelation ($E(\varepsilon_t \varepsilon_\nu) = 0$ for $t \neq \nu$). For ease of implementation, we further assume that the residuals are normally distributed.}
Thus, we can model the likelihood of the outcome (responses to item $1$) at time point $t$ as a normal distribution with variance $\sigma_{y_t}^2$ and mean $\mu_{y_t}$ that takes the $\operatorname{AR}(1)$ structure 
\numberedtabx{
  $\mu_{y_1}  = \alpha$ 
  &\texttt{mu.y[1] <- alpha }\\ 
  &\texttt{for(t in 2:Nt)\{}\\
 $y_{t-1}^{\text{center}}  = y_{t-1}-\alpha$
 &\quad\texttt{
     y.center[t-1] \textless- y[t-1] - alpha}\\
   $\mu_{y_t}  = \alpha + \beta \cdot y_{t-1}^{\text{center}}$ &\quad\texttt{
    mu.y[t] <- alpha + beta * y.center[t-1]}\\
  &\texttt{\},}
}{code:AR1Person1}
which is the same as \Cref{eq:centered_ar1} split in several steps: First the mean at $t=1$ is set to the intercept $\alpha$ (\texttt{alpha}). This needs to happen outside of the loop over the time steps (\texttt{for(t in 2:Nt)\{\}}) since $y_{t-1}$ is not defined for $t=1$. Next, the $\operatorname{AR}(1)$ structure is implemented for the following time steps $t>2$ within the loop, where the centered output $y_{t-1}^{\text{center}}$ (\texttt{y.center[t-1]}) is pre-calculated in an additional step.

Typical priors for the $\operatorname{AR}(1)$ parameters are listed in \Cref{tab:priorsAR1}. 

\begin{table}[H]
    \centering
    \caption{Priors for the parameters of an $\operatorname{AR}(1)$ process.}
    \label{tab:priorsAR1}
    \begin{tabularx}{\textwidth}{lXll}
        \hline
        Parameter & Prior family & Prior & Code\\
        \hline
        Intercept & Normal distribution & $\alpha \sim \mathcal{N}(0,10)$ & \texttt{alpha $\sim$ dnorm(0,0.1)}\\
        & (uninformative) & & \\
        Autocorrelation & uniform distribution & $\beta \sim Unif(-1,1)$ & \texttt{beta $\sim$ dunif(-1,1)}\\
        coefficient & (uninformative) & & \\
        \hline\hline
    \end{tabularx}
\end{table}

The path diagram is displayed in \Cref{fig:02_arma10}.

\input{fig/02_timeseries}

\subsection{Results of Illustration 2}

The intercept, $\alpha$, is estimated at $0.03$ with SD = $0.60$, and the autocorrelation coefficient, $\beta$, at $0.65$ with SD = $0.25$ (\Cref{tab:posterior_ar1_1person_observed}). 

\begin{table}[H]
    \centering
    \caption{Illustration 2: Posterior estimates (posterior mean, standard deviation (SD), and 95\% credible interval defined by the 2.5\% and 97.5\% quantiles) for the $\operatorname{AR}(1)$ model: intercept/baseline value ($\alpha$), autoregressive coefficient ($beta$) and residual variance ($\sigma_{\varepsilon}^2$).}
    \label{tab:posterior_ar1_1person_observed}
    \begin{tabular}{lcccc}
        \hline
        Parameter & Mean & SD & 2.5\% & 97.5\% \\
        \hline
        \( \alpha \) & 0.03 & 0.60 & -0.92 & 1.47  \\
        \( \beta \) & 0.65 & 0.25 & 0.10 & 0.99 \\
        \( \sigma_{\varepsilon}^2 \) & 0.53 & 0.21 & 0.25 & 1.06 \\
        \hline
    \end{tabular}
\end{table}

\subsection{Discussion of Illustration 2}

\begin{figure}[H]
    \centering
    \includegraphics[width=\textwidth]{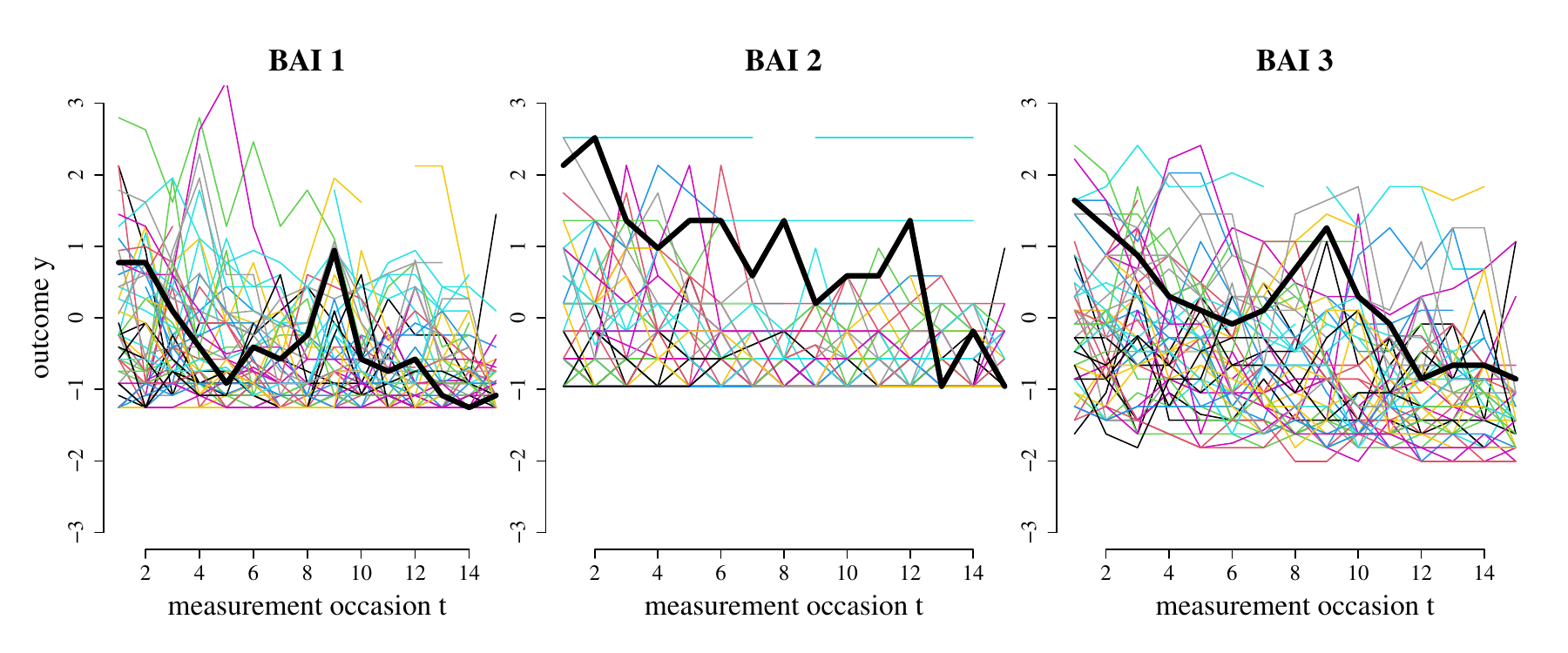}
    \caption{Trajectories of individual scores across 15 measurement occasions for three different BAI items, illustrating heterogeneity across patients.}
    \label{fig:spaghetti_plots}
\end{figure}

\chg{We visualize the data in \Cref{fig:spaghetti_plots} using \textit{spaghetti plots}, which depict patients' response trajectories across measurement occasions. The responses of patient 8 to the first BAI item (highlighted as a thick black line in the first frame) start low, then vary over time and show a decreasing trend. The moderate, statistically significant autocorrelation ($\beta = 0.65$) supports a clear temporal dependence and indicates stationarity. However, a single-item model cannot capture the full latent construct of anxiety, which is later addressed through a measurement model within the DSEM framework.}

%% file: fig/02_timeseries.tex
\begin{figure}[H]
\begin{center}
\resizebox{.9\textwidth}{!}{
\begin{tikzpicture}[>=stealth,semithick]
    \node[lr] (a) at (-1,4)  {(a)};
    
	\node[ov] (y1) at (1,4)  {$y_{1}$};
	\node[ov] (y2) at (5,4)  {$y_{2}$};
	\node[ov] (y3) at (9,4)  {$y_{3}$};
	
	\node[lr] (eps1) at (0,3)      {$\varepsilon_{1}$};
	\node[lr] (eps2) at (4,3)      {$\varepsilon_{2}$};
	\node[lr] (eps3) at (8,3)      {$\varepsilon_{3}$};
	\path[->] 
	(y1)  edge node[above,scale=0.8] {} (y2)
	(y2)  edge node[above,scale=0.8] {} (y3)
	;
	\path[->] 
	(eps1)  edge node[above,scale=0.8] {} (y1)
	(eps2)  edge node[above,scale=0.8] {} (y2)
	(eps3)  edge node[above,scale=0.8] {} (y3)
	;

    \node[lr] (b) at (-1,1)  {(b)};

    \node[ov] (yt) at (1,1)  {$y_{t}$};
	\node[ov] (ytp1) at (5,1)  {$y_{t+1}$};
    \node[lr] (epst) at (0,0)      {$\varepsilon_{t}$};
	\node[lr] (epstp1) at (4,0)      {$\varepsilon_{t+1}$};

    \path[->] 
	(yt)  edge node[above,scale=0.8] {} (ytp1)
	;
	\path[->] 
	(epst)  edge node[above,scale=0.8] {} (yt)
	(epstp1)  edge node[above,scale=0.8] {} (ytp1)
	;

	\end{tikzpicture}
}
\end{center}
\caption{Path diagram for the AR(1) time series model for a single person and a single item $y$. (a) for $3$ explicit time points; (b) for $t\in\{1,...,N_t\}$ time points.}
\label{fig:02_arma10}
\end{figure}

%% file: 03_mlm.tex
\section{Multilevel Modeling (MLM)}
\label{sec:mlm}

In the previous section, we examined the development of a single patient over time \chg{to provide a conceptual baseline for time series modeling}. We now extend this approach to data collected at multiple time points from multiple patients, building what is known as a longitudinal model \cite{Roy2013}. This approach allows us to capture diverse trajectories across patients, making it possible to study inter-individual differences in a structured way. A suitable framework for this purpose is multilevel (hierarchical) modeling \cite{GelmanHill2006}.

\chg{To illustrate the modeling procedure step by step, we begin with an example in which the outcome is directly observable. In the next section, we extend this framework to include latent variables, building a full DSEM}. Here, we focus on constructing a classical multilevel model for longitudinal data. In this framework, the outcome $y_{i t}$ depends on the patients $i\in(1,..,N)$ and time points $t\in(1,..,N_t)$. This outcome can be decomposed into a person-specific between-level model $y_{2i}$, which is constant over time, a time-specific between-level model $y_{3t}$, which is the same for all patients, and a within-level model $y_{1t}$, which may change over time and vary between different patients:
\begin{equation}
\label{output_person_time_decomposition}
    y_{i t}=y_{1it}+y_{2i}+y_{3t}.
\end{equation}
Let us make this more concrete by considering the development of an anxiety item over time and which factors may influence each component:
\begin{itemize}
    \item $y_{2i}$: personal characteristics such as baseline anxiety, gender, or genetic predispositions. These are the person-specific characteristics in the developmental trajectories that remain constant over time or do not depend on it.
    \item $y_{3t}$: the general effectiveness of the intervention over time for anyone participating, which may depend on factors like the experience of the psychotherapist (if the same across patients)
    This component represents the time-specific effect, capturing the average time trend that is independent of the patients and does not convey any information about inter-individual differences in the growth process.
    \item $y_{1it}$: the personal development process which accounts for changes that patients go through during the intervention process, such as responsiveness to the treatment or understanding of the intervention components. These are the person-specific and time-specific aspects.
\end{itemize}

Each part of the model may include random effects, which are captured through residual variances and their correlations. Time-specific deviations are included in $\varepsilon_{3t}$ at the higher-level model $y_{3zi}$, while person-specific deviations are included in $\varepsilon_{2i}$ at the between-level model $y_{2i}$. The errors $\varepsilon_{3t}$/$\varepsilon_{2i}$ may contain both a random intercept $\varepsilon_{3t1}$/$\varepsilon_{2i1}$ and a random slope $\varepsilon_{3t2}$/$\varepsilon_{2i2}$.

\chg{Multilevel parameters provide considerable flexibility, allowing the model to capture heterogeneity in the data. However, this flexibility comes at a cost: increasing the number of parameters consumes degrees of freedom and may reduce interpretability and estimation stability. Therefore, carefully selecting and imposing appropriate model constraints is essential to balance flexibility and parsimony. The following illustration exemplifies how such constraints can be chosen in practice.}

\subsection{Data and Goal of Illustration 3}

Our goal is to examine potential variations across patients and over time by modeling individual trajectories, \chg{addressing H2}. Specifically, we aim to evaluate how much flexibility the model needs for accurately capturing the observed heterogeneity in the responses.
\chg{We analyze item 1 of the BAI scale from $N=57$ patients measured across $N_t=15$ time points, considering both two-level and cross-classified models and assuming an $\operatorname{AR}(1)$ structure for temporal dependence. Depending on parameter estimates, we will assess whether unique baseline levels, patient- or time-specific growth patterns, and correlations among them are necessary, or if some constraints can be applied.}

\subsection{Two-Level Model Implementation with Person-Specific Random Intercept and Slope}

In the two-level model examining outcomes $y_{2i}$ and $y_{1it}$, we incorporate person-specific random intercepts and slopes to capture individual differences in baseline values and growth rates. The random intercept $\varepsilon_{2i1}$ (\texttt{eps2[i,1]}) allows for each person to start at a different starting value. The random slope $\varepsilon_{2i2}$ (\texttt{eps2[i,2]}) represents a constant rate of development (the same at each time step) which may vary in effect direction and steepness between patients. They are both part of the person-specific between-level $2$. We model them with a multivariate normal distribution
\numberedtabx{
\( \boldsymbol{\varepsilon}_{2i} \sim \mathcal{MVN}(\mathbf{\mu}_{\varepsilon_{2}}, \boldsymbol{\Sigma}^2_{\varepsilon_2}) \) 
    &\texttt{eps2[i, 1:2] $\sim$ dmnorm(mu.eps2[1:2],}\\
    &\hspace{3cm}\texttt{psi.eps2[1:2, 1:2])},\\
}{code:eps2Distribution}
where \texttt{eps2[i,1:2]} contains the intercept as the first entry and the slope as the second entry. 

The within-level $1$ is determined by these person-specific random effects and the time-specific $\operatorname{AR}(1)$ process. As previously the outcome $y_{1it}$, the responses of patient $i$ at time $t$ to item $1$, is modeled by a normal distribution,
where the prior for the residual precision $\sigma_{y}^{-2}$ (\texttt{psi.y}) is given in \Cref{tab:priorsCFA}. The person- and time-dependent mean $\mu_{y_{it}}$ (\texttt{mu.y[i, t]}) takes the structure of an $\operatorname{AR}(1)$ process with adapted parameters. In comparison to the $\operatorname{AR}(1)$ process as it was introduced in \codeexamplecref{code:AR1Person1} for one patient in the previous section, $\alpha$ changes to 
\numberedtabx{
\( \alpha_{\text{new,i}} = \alpha+\varepsilon_{2i1} \) 
    &\texttt{alpha.new[i] <- alpha + eps2[i, 1]},\\
}{code:alphaNew}
incorporating the random intercept $\varepsilon_{2i1}$ (\texttt{eps2[i, 1]}). In addition, $\beta$ changes to 
\numberedtabx{
\( \beta_{\text{new,i}} = \beta+\varepsilon_{2i2} \) 
    &\texttt{beta.new[i] <- beta + eps2[i, 2]},\\
}{code:betaNew}
including the random slope $\varepsilon_{2i2}$ (\texttt{eps2[i, 2]}).\footnote{\chg{If the variance of the random slope $\varepsilon_{2i2}$ is large for many individuals, it can lead to non-stationarity, as $\beta_{\text{new},i}$ may exceed 1.}}
Then, we can implement the $\operatorname{AR}(1)$ mean structure in the same way as in \codeexamplecref{code:AR1Person1} with an additional index for patient $i$, using the new time series parameters.
The priors for the $\operatorname{AR}(1)$ parameters $\alpha$ (\texttt{alpha}) and $\beta$ (\texttt{beta}) are specified in \Cref{tab:priorsAR1}, the ones for the precision of the random effects $\boldsymbol{\Sigma}^{-2}_{\varepsilon_2}$ in \Cref{tab:priorsRandomEffects}. The path diagram of the model is displayed in \Cref{fig:03_ar1_Npersons}.

\begin{table}[H]
    \centering
    \caption{Priors for random effects in multilevel modeling.}
    \label{tab:priorsRandomEffects}
    \begin{tabularx}{\textwidth}{lllX}
        \hline
        Parameter & Prior family & Prior & Code\\
        \hline
        Precision & Whishart distribution & $\boldsymbol{\Sigma}^{-2}_{\varepsilon_2} \sim Wish(\boldsymbol\Sigma_0,2)$ & \texttt{psi.eps2[1:2,1:2]}\\
        person-specific & (uninformative) & with hyperprior $\Sigma_0$ & \quad\texttt{$\sim$ dwish(psi0, 2)}\\
        random effects & & & \texttt{psi0=diag(2)}\\
        Precision & Gamma distribution & $\sigma^{-2}_{\varepsilon_3} \sim \Gamma(1,1)$ & \texttt{psi.eps3}\\
        time-specific & (uninformative) & & \quad\texttt{$\sim$ dgamma(1,1)}\\
        random slope & & & \\
        \hline\hline
    \end{tabularx}
\end{table}

\input{fig/03_mlm}

\subsection{Cross-Classified Model Implementation Including Time-Specific Random Slope}

When extending the model to the cross-classified version, we add a time-specific random slope with normal distribution
\numberedtabx{
&\texttt{for(t in 1:Nt)\{}\\
\( \varepsilon_{3t} \sim \mathcal{N}(0, \sigma^2_{\varepsilon_{3}}) \) 
    &\quad\texttt{epsilon3[t] $\sim$ dnorm(0, psi.eps3)}\\
    &\texttt{\}},
}{code:epsilon3Distribution}
which is added to the beta parameter in \codeexamplecref{code:betaNew}
\numberedtabx{
\( \beta_{\text{new},it} = \beta+\varepsilon_{2i2} + \varepsilon_{3t}\) 
    &\texttt{beta.new[i,t] <- beta+eps2[i,2]+eps3[t]},
}{code:betaNewCrossClassified}
allowing for different slopes between different time points for the same patient.

\subsection{Results of Illustration 3}

The $\operatorname{AR}(1)$ parameters, the person-specific random effect variances, and their correlation have similar estimates in both models (\Cref{tab:posterior_mlm}). The additional time-specific random slope of the cross-sectional model has a significant estimated variance of $0.20$ (SD = $0.09$).

\begin{table}[H]
    \centering
    \caption{Illustration 3: Posterior estimates (posterior mean, standard deviation (SD), and 95\% credible interval defined by the 2.5\% and 97.5\% quantiles) for the 2-level and the cross-classified model analyzing responses to a single observable item: parameters for the $\operatorname{AR}(1)$ process (intercept $\alpha$ and autoregressive coefficient $\beta$), covariance of person-specific random effects ($\boldsymbol{\Sigma}^2_{\varepsilon_2}$), variance of the time-specific random slope ($\sigma^2_{\varepsilon_{3}}$),  and residual variance ($\sigma_{y}^2$).}
    \label{tab:posterior_mlm}
    \begin{tabular}{lcccc|cccc}
        \hline
        \multirow{2}{*}{Parameter} & \multicolumn{4}{c|}{2-level model} & \multicolumn{4}{c}{cross-classified model} \\
         & Mean & SD & 2.5\% & 97.5\% & Mean & SD & 2.5\% & 97.5\% \\
        \hline
        $\alpha$ & -0.42 & 0.09 & -0.58 & -0.24 
        & -0.41 & 0.09 & -0.58 & -0.23 \\
        $\beta$  & 0.45 & 0.06 & 0.32 & 0.57
        & 0.44 & 0.14 & 0.16 & 0.70 \\
        $\boldsymbol{\Sigma}^2_{\varepsilon_2}[1,1]$ & 0.33 & 0.08 & 0.20 & 0.52
        & 0.35 & 0.09 & 0.21 & 0.55 \\
        $\boldsymbol{\Sigma}^2_{\varepsilon_2}[1,2]$ & 0.03 & 0.04 &  -0.04 & 0.12
        & 0.03 & 0.04 & -0.05 & 0.11 \\
        $\boldsymbol{\Sigma}^2_{\varepsilon_2}[2,2]$ & 0.11 & 0.03 & 0.06 & 0.18
        & 0.11 & 0.03 & 0.06 & 0.18 \\
        $\sigma^2_{\varepsilon_{3}}$ & - & - & - &  -
        & 0.20 & 0.09 & 0.09 & 0.41 \\
        $\sigma_{y}^2$  & 0.28 & 0.02 & 0.25 & 0.31
        & 0.27 & 0.01 & 0.24 & 0.30 \\
        \hline
    \end{tabular}
\end{table}

\subsection{Discussion of Illustration 3}

\chg{The \textit{spaghetti plot} in the first frame of \Cref{fig:spaghetti_plots} reveals some heterogeneity, with differences more pronounced in baseline levels than in temporal trends. Model estimates confirm this, with moderate variance in random intercepts and small but significant variance in random slopes. Low residual variance in the two-level model suggests that within-individual fluctuations are well captured, indicating a good overall fit. Given the near-zero correlation between random effects, treating intercepts and slopes as independent (\codeexamplecref{code:eps2Distribution}) may provide a more parsimonious alternative. In the cross-classified model, person-specific random effects remain consistent with the two-level model, demonstrating robustness. The significant moderate variance of the additional time-specific random slope indicates notable time-specific fluctuations beyond patient-level trends. Overall, these results support H2: anxiety development exhibits meaningful interindividual differences in both baseline levels and trajectories over time.}

%% file: fig/03_mlm.tex
\begin{figure}[H]
\begin{center}
\resizebox{.8\textwidth}{!}{
\begin{tikzpicture}[>=stealth,semithick]
	\node[ov] (y1) at (1,8)  {$y_{it}$};
	\node[ov,scale=0.8] (y2) at (5,8)  {$y_{i(t+1)}$};
	
	\node[lr] (eps1) at (0,7)      {$\varepsilon_{it}$};
	\node[lr, scale=0.8] (eps2) at (4,7)      {$\varepsilon_{i(t+1)}$};

	\path[->] 
	(y1)  edge node[scale=1] {$\bullet$} (y2)
	;
    \node[re] (beta1) at (3,7.5)     {$\beta_{i}$};

	\path[->] 
	(eps1)  edge node[above,scale=0.8] {} (y1)
	(eps2)  edge node[above,scale=0.8] {} (y2)
	;

    \node[re] (re1) at (1,6.5)     {$\alpha_{i}$};
    \node[re] (re2) at (5,6.5)     {$\alpha_{i}$};
    
    \path[->] 
	(re1)  edge node[scale=1] {$\bullet$} (y1)
	(re2)  edge node[scale=1] {$\bullet$} (y2)
	;

	\draw[dashed] (-2,6) -- (8,6);
	\node (text1) at (7,7) {within};
	\node (text1) at (7,5) {between};
	
	\node[re] (u0) at (2,5)      {$\alpha_{i}$};
   \node[re] (u1) at (4,5)      {$\beta_{i}$};

   \path[<->] 
	(u0)  edge [bend right=50,looseness=.4] (u1)
    ;
  \end{tikzpicture}
}
\end{center}
\caption{Path diagram of the multilevel model for a single observed item $y$ using an AR(1) structure with random intercept ($\alpha_i$) and slope ($\beta_i$) across $t\in\{1,...,N_t\}$ time points and $i\in\{1,..., N\}$ individuals.}
\label{fig:03_ar1_Npersons}
\end{figure}
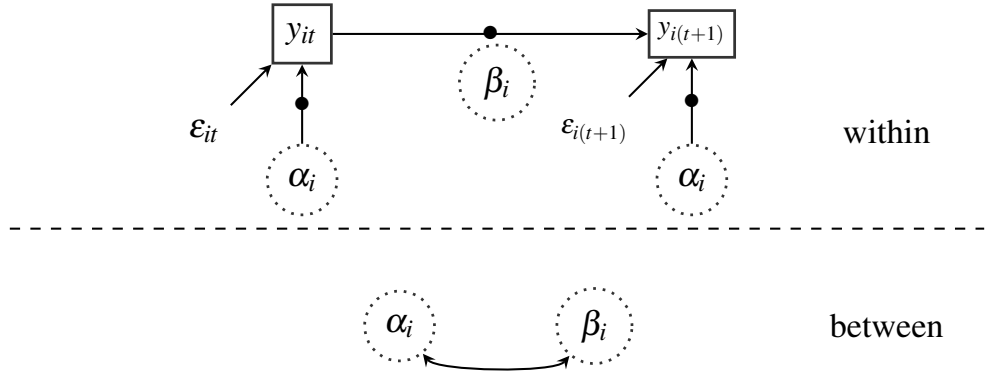

%% file: 04_dsem.tex
\section{Dynamical Structural Equation Modeling (DSEM)}

In this section, we extend the time series modeling approach for multiple patients within the multilevel framework by incorporating latent variable modeling, specifically the CFA models introduced in the first section. This integration creates a dynamical structural equation modeling (DSEM) framework \cite{Asparouhov2018}, allowing us to trace individual trajectories of latent variables over time while accounting for measurement error.

\chg{Instead of} the directly observable outcome $y_{it}$, we model the latent factor $\eta_{it}$. Analogous to the decomposition of $y_{it}$ in \Cref{output_person_time_decomposition}, this latent factor $\eta_{it}$ is divided into three components: a person-specific factor $\eta_{2i}$ that captures temporally stable inter-individual differences; a time-specific factor $\eta_{3t}$, representing systematic variations across time points; and a within-level factor $\eta_{1it}$, which captures unique, moment-to-moment intra-individual fluctuations. 

\chg{The general mathematical formulation of DSEM is provided in Appendix~\ref{App:dsem}. In the following illustration, we focus on an $\operatorname{AR}(1)$ process for one latent factor with random effects. The corresponding link between this specific model and the general DSEM formulation is also detailed in Appendix~\ref{app:dsem_illu4}.}

\subsection{Data and Goal of Illustration 4}

\chg{We extend the multilevel $\operatorname{AR}(1)$ model from the previous section by including a latent factor representing anxiety based on all three BAI items. Following the previous illustration, we consider a model with only person-specific effects, as well as an extended version that additionally includes a time-specific random intercept $\zeta_{3t}$. This modeling strategy allows us to examine interindividual differences in anxiety trajectories and to assess whether temporal dependencies in anxiety change over time, thereby addressing H2 and H3.}

\subsection{Implementation of One-Factor DSEM}

The one-factor model is an easy extension of the two-level model from the last section. We simply replace the distribution of the directly observed outcome $y_{1it}$ (\texttt{y[i,t]}) with the distribution of latent factor 
\numberedtabx{
\( \eta_{it} \sim \mathcal{N}(\mu_{\eta_{it}}, \sigma_{\eta}^2) \) 
    &\quad\texttt{eta[i,t] $\sim$ dnorm(mu.eta[i,t], psi.eta)}.\\
}{code:DSEM1factorDistribution}
Then we apply exactly the  same structure to $\mu_{\eta_{it}}$ (\texttt{mu.eta[i,t]}) as to $\mu_{y_{1it}}$ (\texttt{mu.y[i,t]}) in the previous example for an $\operatorname{AR}(1)$ time structure, changing only the variable names.
The random intercept and slope $\zeta_{2i}$ of the latent variable are defined in the same way as random intercept and slope $\varepsilon_{2i}$ of the observed variable in \crefrange{code:eps2Distribution}{code:betaNew}.

The factor model is directly taken from the first section, with the assumption that it remains the same across all patients and time points, so its definition lies outside the multilevel framework.  Therefore, the measurement model of the observed outcome $y_{itj}$ of indicator $j$ and patient $i$ at time $t$ is the same as in the time series model for one patient  with an additional patient index $i \in (1,...,N)$:
\numberedtabx{
$y_{itj}\sim \mathcal{N}(\mu_{y_{itj}},\sigma^2_j)$
    &\quad\quad\quad\texttt{y[i,t,j] ~$\sim$ dnorm(mu.y[i,t,j],}\\
    &\hspace{5.3cm}\texttt{psi.y[j])}\\
    $\mu_{y_{it1}}=\eta_{it}$
    &\quad\quad\texttt{mu.y[i,t,1] <-  eta[i,t]}\\
$\mu_{y_{it2}}=\nu_1+\lambda_1\cdot\eta_{it}$
    &\quad\quad\texttt{mu.y[i,t,2] <- nu.y[1]+lambda.y[1]*eta[i,t]}\\
$\mu_{y_{it3}}=\nu_2+\lambda_2\cdot\eta_{it}$
    &\quad\quad\texttt{mu.y[i,t,3] <- nu.y[2]+lambda.y[2]*eta[i,t]}\\
}
The path diagram is displayed in \Cref{fig:04_ar1_Npersons_latent}. The priors of the single factor's mean and precision and the two free factor loadings and intercepts are the same as in \Cref{tab:priorsCFA} (adapting the ranges of the for-loops accordingly). The extension for the time-specific random slope is identical to the previous implementation, switching out $\varepsilon_{3t}$ with $\zeta_{3t}$.

\input{fig/04_dsem}

\subsection{Results of Illustration 4}

The two-level model estimates an intercept of $-0.32$ ($SD=0.10$) and an autoregressive coefficient of $0.70$ ($SD=0.07$; \Cref{tab:posterior_ar1_allpersons_1factor}). The within-level variance of the anxiety factor is small ($0.07$, $SD=0.01$). 

\begin{table}[H]
\centering
    \caption{Illustration 4: Posterior estimates (posterior mean, standard deviation (SD), and 95\% credible interval defined by the 2.5\% and 97.5\% quantiles) for the $\operatorname{AR}(1)$ latent variable model: intercept/baseline value ($\alpha$), autoregressive coefficient ($\beta$), factor variance ($\sigma_{\eta}$), factor loadings ($\lambda$), factor intercepts ($\nu$), covariance of person-specific random effects ($\boldsymbol{\Sigma}^2_{\zeta_2}$), variance of the time-specific random effect ($\sigma^2_{\zeta_{3}}$) and residual variance ($\sigma_{y_j}^2$) for each indicator $j$.}
    \label{tab:posterior_ar1_allpersons_1factor}
    \begin{tabular}{lcccc|cccc}
        \hline
        \multirow{2}{*}{Parameter} & \multicolumn{4}{c|}{2-level model} & \multicolumn{4}{c}{cross-classified model} \\
         & Mean & SD & 2.5\% & 97.5\% & Mean & SD & 2.5\% & 97.5\% \\
        \hline
        $\alpha$ & -0.29 & 0.10  & -0.49  & -0.10  
        & -0.39 & 0.10  & -0.58 & -0.19  \\
        $\beta$ & 0.71 & 0.07 & 0.58 & 0.84 
        &  0.65 & 0.15 & 0.35 & 0.92  \\
        $\sigma_{\eta}^2$ &  0.07 & 0.01 & 0.06 & 0.09
        &  0.07 & 0.01 & 0.05 & 0.08  \\
        $\lambda_1$ & 0.89 & 0.05 & 0.79  & 0.99
        &  0.90 & 0.05 & 0.80 &  1.00 \\
        $\lambda_2$ & 1.36 & 0.07 & 1.23 & 1.51
        & 1.34 & 0.07 & 1.21 & 1.47 \\
        $\nu_1$ &  0.05 & 0.04 & -0.03 & 0.12
        &  0.05 & 0.04 & -0.02 & 0.12  \\
        $\nu_2$ &  0.10 & 0.05 & 0.01 & 0.20
        &  0.09 & 0.05 & 0.01 & 0.1  \\
        $\boldsymbol{\Sigma}^2_{\zeta_2}[1,1]$ &  0.40 & 0.10 & 0.24 & 0.63
        &  0.42 & 0.11  &  0.24 & 0.68  \\
        $\boldsymbol{\Sigma}^2_{\zeta_2}[1,2]$ &  0.10 & 0.04 & 0.03 & 0.19
        &  0.12 & 0.05  &  0.04 & 0.23 \\
        $\boldsymbol{\Sigma}^2_{\zeta_2}[2,2]$ & 0.09 & 0.03 & 0.05 & 0.16
        &  0.12 & 0.04 & 0.06 & 0.21  \\
        $\sigma^2_{\zeta_{3}}$ & - & - & - &  -
        & 0.22 & 0.10 & 0.10 & 0.49 \\
        $\sigma_{y_1}^2$ & 0.28 & 0.02 & 0.25 & 0.33
        &  0.28 & 0.02 & 0.25 & 0.32  \\
        $\sigma_{y_2}^2$ & 0.33 & 0.02 & 0.29 & 0.38
        &  0.33 & 0.02 & 0.28 & 0.38  \\
        $\sigma_{y_3}^2$ & 0.26 & 0.03 & 0.19 & 0.32
        &  0.27 & 0.03 & 0.21 & 0.34  \\
        \hline
    \end{tabular}
\end{table}

\chg{Person-specific random effects show a variance of $0.40$ ($SD=0.10$) for the intercept and $0.09$ ($SD=0.03$) for the slope, with a correlation of $0.52$.} 
\chg{The cross-classified model produces results very similar to the two-level model. The estimated time-specific random variance is $0.22$ ($SD=0.10$).}

\subsection{Discussion of Illustration 4}

\chg{Person-specific random effect variances estimated with the DSEM are similar to those obtained from the MLM with observed variables. These results confirm the presence of interindividual differences as stated in H2. As illustrated in \Cref{fig:spaghetti_plots} and reflected by the substantial random intercept variance, these differences are particularly pronounced in baseline anxiety levels.}

\begin{figure}[ht!]
    \centering
    \includegraphics[scale=0.75]{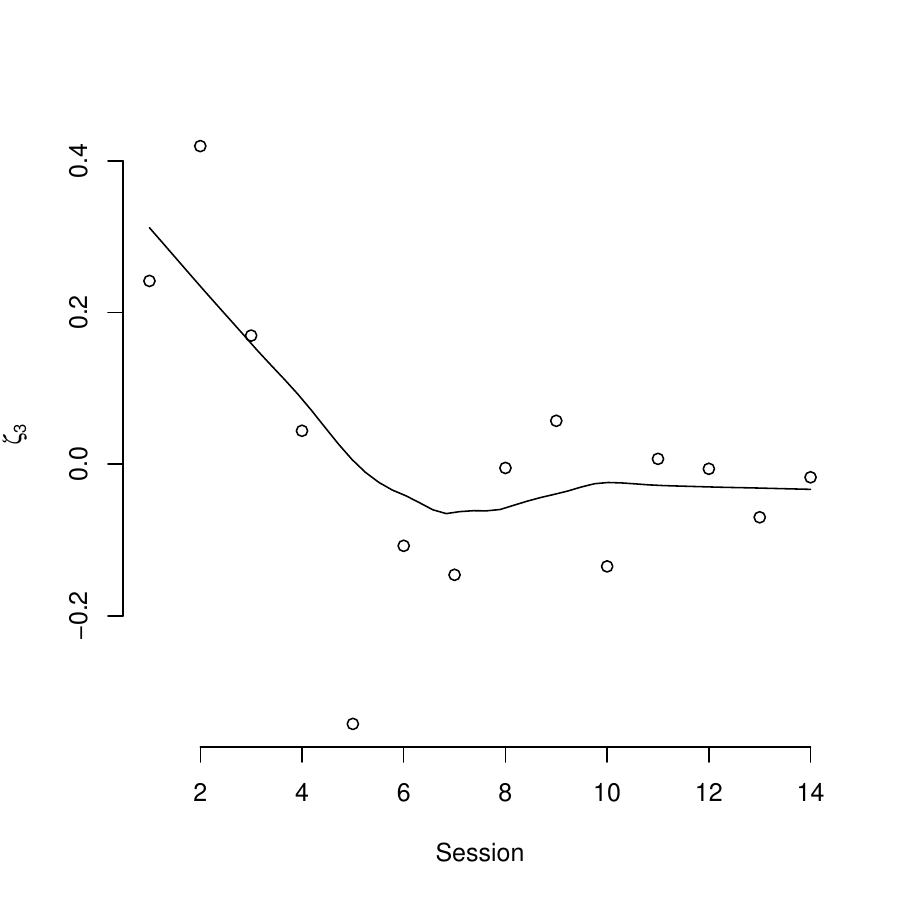}
    \caption{Estimates scores for for the time-specific random slope $\zeta_3$ across sessions. The solid line indicates a loess approximation to highlight the development.
    }
    \label{fig:timeplot}
\end{figure}

\chg{The high autocorrelation supports the dynamic evolution of anxiety as formulated in H3. The significant variance of the time-specific random effect further confirms temporal variation in dependencies. Figure~\ref{fig:timeplot} shows that autocorrelation is highest at the beginning, declines until session 6, and then stabilizes at a lower level, reflecting increased within-person variability in anxiety. In Illustration 5, we will examine how these dynamics manifest at the level of individual patients using DLCSEM.}

\subsection{Learnings Regarding the Procedure}

\chg{This step concludes the second part of our modeling guidelines. At this stage, researchers should have established several key aspects of the model. First, they should have evaluated whether the factor structure is stable across time or shows meaningful changes (see CFA section). Second, they should have identified which variables exhibit temporal dependencies and determined an appropriate time series representation, including the type of process, the number of lags required, and whether the process can be considered stationary (see Time Series section). Third, they should have assessed the degree of heterogeneity in the data, both between individuals and across time points, and decided which person-specific and time-specific random effects, as well as their associations, need to be modeled and which can be constrained (see Multilevel Model section).}

\chg{The overall goal is to identify a level of model complexity that adequately captures the essential features of the data while maintaining stable and interpretable parameter estimates. Particular attention should be paid to the specification of random effects. Incorrectly assuming homogeneity when meaningful heterogeneity is present can bias parameter estimates and lead to misleading conclusions about the underlying process. Conversely, introducing excessive flexibility through unnecessary random effects can result in overfitting, produce artifacts that reflect noise rather than substantive patterns, and complicate interpretation by obscuring common trends.}

\chg{Theoretical considerations play a crucial role in guiding these decisions and help define a model that is sufficiently parsimonious while remaining substantively meaningful. At the same time, empirical support must be evaluated. This includes examining model convergence, the magnitude and uncertainty of parameter estimates, as well as overall model fit. However, it should be noted that clear and universally accepted model fit criteria for DSEM are still an area of ongoing methodological development.}

%% file: fig/04_dsem.tex
\begin{figure}[ht!]
\begin{center}
\resizebox{.7\textwidth}{!}{
\begin{tikzpicture}[>=stealth,semithick]
	
	\node[ov] (y11) at (0,1)  {$y_{it1}$};
	\node[ov] (y12) at (1,1)  {$y_{it2}$};
	\node[ov] (y13) at (2,1)  {$y_{it3}$};
	\node[lv] (eta1) at (1,0)  {$\eta_{it}$};
	\node[lr] (zeta1) at (0,-1)  {$\zeta_{it}$};

	\node[ov,scale=0.8] (y21) at (3.5,1)  {$y_{i(t+1)1}$};
	\node[ov,scale=0.8] (y22) at (5,1)  {$y_{i(t+1)2}$};
	\node[ov,scale=0.8] (y23) at (6.5,1)  {$y_{i(t+1)3}$};
	\node[lv,scale=0.6] (eta2) at (5,0)  {$\eta_{i(t+1)}$};
	\node[lr,scale=0.8] (zeta2) at (4,-1)  {$\zeta_{i(t+1)}$};
	
	\path[->] 
	(eta1)  edge node[above,scale=0.8] {} (y11)
	(eta1)  edge node[above,scale=0.8] {} (y12)
	(eta1)  edge node[above,scale=0.8] {} (y13)
    (eta2)  edge node[above,scale=0.8] {} (y21)
	(eta2)  edge node[above,scale=0.8] {} (y22)
	(eta2)  edge node[above,scale=0.8] {} (y23)
    ;
	\path[->] 
	(eta1)  edge node[scale=1] {$\bullet$}  (eta2)
	;

    \node[re] (re1a) at (3,-.5)     {$\beta_{is}$};
    
	\path[->] 
	(zeta1)  edge node[above,scale=0.8] {} (eta1)
	(zeta2)  edge node[above,scale=0.8] {} (eta2)
	;

    \node[re] (re1) at (1,-1.5)     {$\alpha_{i}$};
    \node[re] (re2) at (5,-1.5)     {$\alpha_{i}$};
    
    \path[->] 
	(re1)  edge node[scale=1] {$\bullet$} (eta1)
	(re2)  edge node[scale=1] {$\bullet$} (eta2)
	;

	\draw[dashed] (-2,-2) -- (8,-2);
	\node (text1) at (7,-1) {within};
	\node (text1) at (7,-3) {between};
	
	\node[re] (u0) at (2,-3)      {$\alpha_{i}$};
   \node[re] (u1) at (4,-3)      {$\beta_{i}$};

   \path[<->] 
	(u0)  edge [bend right=50,looseness=.4] (u1)
    ;

	\end{tikzpicture}
}
\end{center}
\caption{Path diagram of the DSEM for the latent factor $\eta$ loading on three indicators $y$ using an AR(1) structure with random intercept ($\alpha_i$) and slope ($\beta_i$) across $t\in\{1,...,N_t\}$ time points and $i\in\{1,...,N\}$ individuals.}
\label{fig:04_ar1_Npersons_latent}
\end{figure}

%% file: 05_hmsm.tex
\newpage
\section{Hidden Markov Switching Models (HMSM)}
\label{sec:hmm}

\chg{In this section, we introduce another modeling layer: the Hidden Markov Switching Model (HMSM). To reveal its purpose, we begin by untangling the ideas embedded in its name.}

\chg{\textit{Hidden} refers to an additional latent variable, the latent state, which represents unobserved internal conditions that influence how other variables change over time. This latent state is similar to a latent factor in that it shapes patterns in the observed data. However, it plays a different role: rather than representing a stable underlying trait, it governs the evolution of several observable or latent traits. For example, an internal stressor may affect physiological measures such as cortisol levels and heart rate, while also shaping psychological constructs such as life satisfaction and emotional stability. We can thus think of the latent state as a background mode of functioning: when this mode changes, the joint behavior of multiple variables changes accordingly.}
\chg{Historically, this idea stems from State–Space Models (SSMs), which describe systems in terms of a hidden state governing observable time-series data. The classical formulation, introduced by \citet{Kalman1960} as a filtering technique in control engineering, assumes that this latent state evolves on a continuous scale \citep[for an overview, see][]{Durbin2012}. In this work, however, we focus on SSMs with a finite number of discrete latent states. Such Hidden Markov Models \citet[HMMs][]{Hamilton1989} model distinct internal conditions, corresponding to a latent class, such as different stress modes (e.g., high stress, moderate stress, calm). 
}

\chg{\textit{Markov} refers to a property of the latent state concerning its temporal structure. Similar to lagged dependencies in time-series models, the latent state may depend on a limited number of its past values. This is known as the Markov property of order $p$, which means that the state at time $t$ depends only on the previous $p$ states, rather than on the full history of earlier states. Here, we assume a first-order Markov property, corresponding to a short-term memory in which the current state depends solely on the immediately preceding state (similar to an $\operatorname{AR}(1)$ process).}

\chg{\textit{Switching} refers to the mechanism we aim to capture: individuals are allowed to transition between latent states over time, enabling us to detect changes in their underlying condition. For example, we may be interested in observing whether a person moves from a high-stress latent class to a moderate-stress or calm class, and how such shifts are reflected in physiological measures (e.g., cortisol, heart rate) and psychological constructs (e.g., life satisfaction, emotional stability). Modeling these transitions explicitly allows us to understand not only the states themselves but also the dynamics of how individuals evolve from one internal condition to another.}

\chg{To model such individual-level state switches, we make use of another class of models known as regime-switching models. These models were introduced by \citet{Hamilton1989} to capture abrupt structural shifts in economic time series, including financial data \cite{Hamilton2018}.}\footnote{\chg{For a detailed introduction to SSMs with regime-switching, including a Bayesian implementation, please refer to \citet{Chang1999}.}}

\chg{In our application to patient data, each individual occupies exactly one latent state at each time point. The regime-switching model then governs how the individual may transition between these states at future time points. To accommodate this, we define the transition $\text{P}$, whose entries}
\[
    \text{P}_{ss'} = P(S_t = s \mid S_{t-1} = s')
\]
\chg{specify the likelihood of moving from state $s'$ at time $t-1$ to state $s$ at time $t$. The full matrix has as many rows and columns as there are states. Diagonal entries contain the probabilities of remaining in the same state, while the off-diagonal entries specify the probabilities of switching states.}

\chg{For two latent states $1$ and $2$, the complete transition matrix is:}
\begin{equation*}
    \begin{bmatrix}
        P(S_{t} = 1 \mid S_{t-1} = 1) & P(S_{t} = 2 \mid S_{t-1} = 1)  \\
        P(S_{t} = 1 \mid S_{t-1} = 2)  & P(S_{t} = 2 \mid S_{t-1} = 2) 
    \end{bmatrix}
    =:\begin{bmatrix}
        \text{P}_{11} & \text{P}_{12} \\
        \text{P}_{21}  & \text{P}_{22}
    \end{bmatrix}.
\end{equation*}
\chg{Out of the four probabilities, we only need to model two because the other two can be calculated via $\text{P}_{j2} = 1 - \text{P}_{j1}$ for $j\in\{1,2\}$. The modeling task therefore reduces to specifying $\text{P}_{11}$ and $\text{P}_{21}$. These entries can be defined as simple numeric probabilities or as more flexible functions, such as logistic models that allow time- or person-specific transition probabilities (or multinomial models for extensions with more than two states). Next, we describe the implementation based on a logit model.}

\subsection{Implementation of the Transition Model}

The probability $\texttt{P11[i,t]}$ for person $i$ of remaining in state $S_1$ at time \( t \), conditional on being in state $S_1$ at time \( t-1 \), is modeled as:
\numberedtabx{
$\text{logit}(\text{P}_{11it}) = b_{21}$ & \texttt{logit(P1[i,t,1]) <- b2[1]} \\
\hspace{2.2cm}$+ b_{22} \cdot \eta_{1i(t-1)}$ & \hspace{4cm}\texttt{+ b2[2] * eta.S1[i,t-1]}\\
$\text{P}_{12it} = 1 - \text{P}_{11it}$ & \texttt{P1[i,t,2] <- 1 - P1[i,t,1]}
}{code:logistic_predictor}
with a censored prior for $b_{22}\sim \mathcal{N}^{+}(0,1)$, which implied that a transition was more likely if the patient showed lower anxiety scores in the previous session.

Transitions from state $S_2$ are assumed to be rather unlikely, which is defined using a uniform prior:
\numberedtabx{
$\text{P}_{21} \sim Unif(0,.1)$ & \texttt{P2[1] $\sim$ dunif(0,.1)} \\
$\text{P}_{22} = 1 - \text{P}_{21} $ & \texttt{P2[2] <- 1 - P2[1]} \\
}{code:fixed_transition}
As an alternative, a fixed constant such as $\text{P}_{21} =.05$ could be used, which often improves convergence because this simplification stabilizes the sampling.

For all time points \(t > 1\), the transition probability \chg{depends on the state at \(t-1\): if patient \(i\) was in state $S_1$, we use \(\texttt{P}_{1\cdot}\), and if the patient was in state $S_2$, we use \(\texttt{P}_{2\cdot}\). Using this transition probability, the state of patient \(i\) at time \(t\) is then sampled from the categorical distribution:}
\numberedtabx{
$S_{it} \sim \text{Categorical}(P_{S_{it}})$ & 
\quad\quad\texttt{S[i,t] \textasciitilde{} dcat(PS[i,t,1:2]),} \\
}{code:state_sampling}
\chg{which returns either 1 or 2 in the case of two states. We assume that all patients start in \(S=1\), reflecting the idea that individuals begin the study still in need of treatment. If heterogeneity in baseline states is expected, this assumption can be relaxed or adapted, for example, by introducing a prior over the initial state probabilities.}

%% file: 06_dlcsem.tex
\section{DLCSEM}

\chg{By incorporating an HMSM into DSEM, we obtain DLCSEM, able to capture categorical shifts over time, represented by time-dependent latent dynamic states.} Its original implementation was provided in \citet{Asparouhov2017}, and it was later extended to include more flexible nonlinear relationships between variables and time-dependent covariates in the HMSM \cite{Kelava2019}. 

Like DSEM, DLCSEM is based on the multilevel decomposition of the outcome defined in \Cref{output_person_time_decomposition}. To incorporate the latent state, we introduce the variable \( S_{it} \) \chg{as described in the previous section. Then, we model the within-level observed outcome vector as \( \left[Y_{1it} \mid S_{it} = s \right] \) and the within latent variables as \( \left[\eta_{1it} \mid S_{it} = s \right] \), indicating that under each state a different process evolves}. 

In its most general form, all parameters on the within-level can be state-specific, leading to completely different models for each state \chg{(see details in Appendix~\ref{app:dlcsem}). However, as argued above, too much flexibility typically results in either data-driven artifacts or non-convergence. Instead, it is meaningful to constrain parameters using theoretical considerations.}

\chg{Next, we present two illustrations of how to formulate a DLCSEM to study different aspects of process data. Illustration 5 extends the DSEM analysis from the previous example by defining states based on patients’ anxiety levels and incorporating inter-individually varying time points of switching. Illustration 6 focuses on structurally distinct states, allowing us to examine changes in the factor structure over time that were anticipated in the CFA analysis.}

\subsection{Data and Goal of Illustration 5}

We extend the previous example on DSEM \chg{to study anxiety progression dynamics manifest at the level of individual patients using DLCSEM.} 
\chg{We extend Illustration 4 by translating hypothesis H4 into the DLCSEM framework:} instead of using a time-specific random slope, we aim at identifying patients with sudden changes in their anxiety levels. This is represented by a shift in the intercept and autocorrelation parameters of the latent factor at some person-specific time point $t$. The resulting model provides a more detailed interpretation compared to the DSEM because it allows for simultaneous person- and time-specific dynamic changes.

\subsection{Implementation of the One-Factor DLCSEM}

We assume the existence of two distinct latent \chg{states with respect to the latent anxiety factor: a non-improved state and an improved state. Individuals may transition from the non-improved state to the improved state over time. We also assume that once in the improved state, the probability of transition back to the non-improved state is low.}

\chg{This translates to a latent dynamic model with separate latent factors, $\eta_{S_1it}$ and $\eta_{S_2it}$, that are modeled as normally distributed with state-specific precisions:}

\numberedtabx{
    $\eta_{S_1it} \sim \mathcal{N}( \mu_{\eta_{S_1it}}, \sigma_{\eta_{S_1}}^2 ) $  
    &\quad\quad\texttt{eta.S1[i,t] $\sim$ dnorm(mu.eta.S1[i,t], }\\ 
    &\hspace{5.3cm}\texttt{psi.eta.S1)}\\
    $\eta_{S_2it} \sim \mathcal{N}( \mu_{\eta_{S_2it}}, \sigma_{\eta_{S_2}}^2 ) $  
    &\quad\quad\texttt{eta.S2[i,t] $\sim$ dnorm(mu.eta.S2[i,t],}\\ 
    &\hspace{5.3cm}\texttt{psi.eta.S2)}\\
}{code:dlc_eta_distr}

with equal variances across states ($\sigma_{\eta_{S_1}}^2 =\sigma_{\eta_{S_2}}^2 $) and with \chg{$\operatorname{AR}(1)$ mean structure}

\numberedtabx{
    $\mu_{\eta_{S_1it}} = \alpha_{\text{new},1,i} +$
    &\quad\quad\texttt{mu.eta.S1[i, t] <- alpha.S1.new[i] + } \\ 
    \hspace{1.4cm}$ \beta_{\text{new},1,i} \eta_{S_1it}^{\text{center}} $  
    &\hspace{0.5cm} \texttt{beta.S1.new[i] * eta.center.S1[i, t-1], }\\ 
    $\mu_{\eta_{S_1it}} = \alpha_{\text{new},2,i} +$  
    &\quad\quad\texttt{mu.eta.S2[i, t] <- alpha.S2.new[i] + } \\ 
    \hspace{1.4cm}$ \beta_{\text{new},2,i} \eta_{S_1it}^{\text{center}}$ &\hspace{0.5cm} \texttt{beta.S2.new[i] * eta.center.S2[i, t-1], }\\
}{code:dlc_mu_distr}

\chg{where $\alpha_{\text{new},1,i}, \beta_{\text{new},1,i}$, and $\beta_{\text{new},2,i}$ follow the same definition as in the two-level model (\Cref{code:alphaNew} and \Cref{code:betaNew}), except that the correlation between the random intercepts is set to zero to reduce complexity.}

\chg{In order to avoid state-switching and in order to provide a confirmatory specification of the states, we use a specific model for $\alpha_{\text{new},2,i}$ that ensures that patients under state $S_2$ exhibit lower anxiety levels than under state $S_1$:}
\numberedtabx{
    $\alpha_{\text{new},2,i} = \alpha_{\text{new},1,i} - \delta_\alpha $  
    &\texttt{alpha.S2.new[i] <- alpha.S1.new[i]-} \\ 
    &\hspace{5.3cm} \texttt{delta.alpha}\\ 
    $\delta_\alpha \sim \mathcal{N}^{+}(0, 1),$  
    &\texttt{delta.alpha $\sim$ dnorm(0,1)}\texttt{$\cdot$I(0,)}\\
}{code:alpha_truncated}
where the positive censoring ($\mathcal{N}^{+}$) enforces our hypothesis. This constraint has an additional advantage, it helps with the problem of label switching \cite{Stephens2000}: Each state is unambiguously defined due to the intercepts that ensure $\alpha_{1S_1}\geq \alpha_{1S_2}$ because $\delta_\alpha\geq 0$. 
\chg{In principle, larger values for $\delta_\alpha$ could be encouraged by choosing a hyperparameter for the mean of the censored normal distribution, for example, coding clinically relevant effect sizes.}
Note that this definition assumes that the random intercept and random slope covariance matrix (for $\zeta_{2i1},\zeta_{2i2}$) are state-invariant, which implies that there are no baseline differences between patients who will undergo a sudden change and those who do not (which seems plausible). 

The observed BAI items ($y_{1 i t}$) are modeled conditional on the latent state $S_{it} = s$. 
\numberedtabx{
$y_{itj}\sim \mathcal{N}(\mu_{y_{itjs}},\sigma^2_j)$
    &\quad\quad\quad\texttt{y[i,t,j] ~$\sim$ dnorm(mu.y[i,t,j,s],}\\
    &\hspace{5.3cm}\texttt{psi.y[j])}\\
}{code:AR1AllPersonsLatentMeasurementModel}
with \chg{a factor model for state $S_1$}
\numberedtabx{
    $\mu_{y_{it1}}=\eta_{S_1it}$
    &\quad\quad\texttt{mu.y[i,t,1,1] <-  eta.S1[i,t]}\\
$\mu_{y_{it2}}=\nu_1+\lambda_1\cdot\eta_{S_1it}$
    &\quad\quad\texttt{mu.y[i,t,2,1] <- nu.y[1]+lambda.y[1]*eta.S1[i,t]}\\
$\mu_{y_{it3}}=\nu_2+\lambda_2\cdot\eta_{S_1it}$
    &\quad\quad\texttt{mu.y[i,t,3,1] <- nu.y[2]+lambda.y[2]*eta.S1[i,t]}\\
}{code:AR1AllPersonsLatentMeasurementModel_1}
and \chg{a separately defined factor model for state $S_2$}
\numberedtabx{
    $\mu_{y_{it1}}=\eta_{S_2it}$
    &\quad\quad\texttt{mu.y[i,t,1,2] <-  eta.S2[i,t]}\\
$\mu_{y_{it2}}=\nu_1+\lambda_1\cdot\eta_{S_2it}$
    &\quad\quad\texttt{mu.y[i,t,2,2] <- nu.y[1]+lambda.y[1]*eta.S2[i,t]}\\
$\mu_{y_{it3}}=\nu_2+\lambda_2\cdot\eta_{S_2it}$
    &\quad\quad\texttt{mu.y[i,t,3,2] <- nu.y[2]+lambda.y[2]*eta.S2[i,t]}\\
}{code:AR1AllPersonsLatentMeasurementModel_2}
The factor model \chg{retains the same structure as in the DSEM, with factor loadings and intercepts constrained to be invariant across states. This allows a consistent interpretation of the latent variable in both states.} 

\chg{The only state-dependent parameters are} the $\operatorname{AR}(1)$ intercept and autocorrelation of the latent factor \chg{to allow the latent factor to evolve differently across states. The HMSM that models transitions between the states is implemented according to \cref{code:logistic_predictor} to \cref{code:state_sampling}.} The switching point is interpreted as a sudden gain\chg{, as it corresponds to the moment when patients show a (sudden) reduction in symptoms.} The path diagram is depicted in \Cref{fig:02_arma11}.

\input{fig/06_dlcsem}

\begin{table}[H]
    \centering
    \caption{Prior distributions for the latent state-switching AR(1) 2-level model parameters. For the priors of the residual precision, the free factors loadings and intercepts, refer to \Cref{tab:priorsCFA}.}
    \label{tab:priorsLatentModel}
    \begin{tabularx}{\textwidth}{p{3cm}Xlp{6cm}}
        \hline
        Parameter & Prior family & Prior & Code\\
        \hline
        Factor precision & Gamma & $\sigma_{\eta_{S_1}}^{-2} \sim \Gamma(1,1)$ & \texttt{psi.eta.S1 $\sim$ dgamma(1, 1)} \\
        state $S_1$ & & & \\
        Random intercept & Gamma & $\sigma^{-2}_{\zeta_{21}} \sim \Gamma(1,1)$ & \texttt{psi.zeta21} $\sim$ dgamma(1, 1)\\
        precision & & & \\
        AR(1) intercept state $S_1$ & Normal & $\alpha_{1S_1} \sim \mathcal{N}(0, 1000)$ & \texttt{alpha.S1 $\sim$ dnorm(0, 0.001)} \\
        %
        AR(1) difference & Censored Normal& $\delta_{\alpha} \sim \mathcal{N}^+\left(0, 1\right)$ & \texttt{delta.alpha $\sim$ dnorm(0, 1) I(0,)} \\
        AR(1) intercept state $S_2$ & Difference & $\alpha_{1S_2} = \alpha_{1S_1} - \delta_{\alpha}$ & \texttt{alpha.S2 <- alpha.S1 - delta.alpha} \\
        AR(1) coefficient state $S_1$& Uniform & $\beta_{S_1} \sim \mathcal{U}(0, 1)$ & \texttt{beta.S1 $\sim$ dunif(0,1)} \\
        AR(1) coefficient state $S_2$& Uniform & $\beta_{S_2} \sim \mathcal{U}(0, 1)$ & \texttt{beta.S2 $\sim$ dunif(0,1)} \\
        Transition & Normal & $b_{21} \sim \mathcal{N}(0, 100)$ & \texttt{b2[1] $\sim$ dnorm(0, 0.01)} \\
        intercept & & & \\
        Transition slope& Censored Normal& $b_{22} \sim \mathcal{N}^{+}(0, 1)$ & \texttt{b2[2] $\sim$ dnorm(0, 1)I(0, )} \\
        \hline\hline
    \end{tabularx}
\end{table}

\subsection{Results of Illustration 5}

The estimated intercepts for the latent anxiety factor indicate two well-separated states. State $S_1$, the non-respondent condition, was characterized by a higher baseline level of anxiety with an intercept of $\alpha_1=-0.0\chg{3}$ ($SD = 0.0\chg{9}$) compared to state $S_2$, the respondent state, with $\alpha_2=-0.6\chg{2}$ ($SD = 0.0\chg{9}$). The autocorrelations were similar but slightly smaller in state $S_1$ with $\beta_1=0.\chg{44}$ ($SD = 0.1\chg{9}$) compared to state $S_2$ with $\beta_2=0.5\chg{1}$ ($SD = 0.1\chg{1}$), implying that the subsequent time points were better to predict after switching.

The transition model with an impact of $\eta_{S_1i(t-1)}$ on the probability to remain in state $S_1$ if patients were in state $S_1$ at $t-1$ was $0.4\chg{5}$ ($SD=0.34$), which implied that lower anxiety scores provided higher probabilities to switch to state $S_2$. 

\chg{Measurement model estimates (factor loadings, intercepts, and residual variances) were similar to the ones reported above. Random effects were slightly different with a random intercept variance of $0.30$ (compared to $0.40$ in Illustration 5) and a random slope variances of $0.16$ (compared to $0.09$).}

\begin{table}[H]
    \centering
   \caption{Illustration 5: Posterior estimates (posterior mean, standard deviation (SD), and 95\% credible interval defined by the 2.5\% and 97.5\% quantiles), $\hat{R}$ statistic and effective sample size ($n_{eff}$) for the state-switching model: transition probability from state $S_2$ in state $S_1$ ($P_{21}$), transition parameters $b_{21}$ and $b_{22}$, state-dependent intercepts ($\alpha_{1S_1}$ and $\alpha_{1S_2}$), difference between them ($\delta_{\alpha}$), state-dependent autoregressive coefficients ($\beta_{S1}$ and $\beta_{S2}$), variances of person-specific random effects ($\boldsymbol{\Sigma}^2_{\zeta_2}$), within level factor residual variance ($\sigma_{\eta}^2$), 
   factor loadings ($\lambda$), factor intercepts ($\nu$) and residual variances ($\sigma_{y_j}^2$) for each indicator $j$.}
    \label{tab:posterior_DLCSEM}
    \begin{tabular}{rrrrrrr}
      \hline
     & mean & sd & 2.5\% & 97.5\% & $\hat{R}$ & $n_{\text{eff}}$ \\ 
      \hline
    $P_{21}$ & 0.02 & 0.01 & 0.00 & 0.04 & 1.04 & 100 \\ 
      $b_{21}$ & 0.95 & 0.30 & 0.37 & 1.54 & 1.03 & 120 \\ 
      $b_{22}$ & 0.45 & 0.34 & 0.02 & 1.26 & 1.00 & 800 \\ 
      $\alpha_{1S_1}$ & -0.03 & 0.09 & -0.21 & 0.15 & 1.01 & 460 \\ 
      $\alpha_{1S_2}$ & -0.62 & 0.09 & -0.79 & -0.45 & 1.01 & 370 \\ 
      $\delta_{\alpha}$ & 0.59 & 0.06 & 0.48 & 0.71 & 1.01 & 290 \\ 
      $\beta_{S1}$ & 0.44 & 0.19 & 0.06 & 0.80 & 1.07 & 51 \\ 
      $\beta_{S2}$ & 0.51 & 0.11 & 0.27 & 0.70 & 1.07 & 69 \\ 
      $\sigma^2_{\zeta_{21}}$ & 0.30 & 0.07 & 0.19 & 0.47 & 1.01 & 260 \\ 
      $\sigma^2_{\zeta_{22}}$ & 0.16 & 0.05 & 0.09 & 0.29 & 1.03 & 95 \\ 
      $\sigma_{\eta}^2$ & 0.05 & 0.01 & 0.04 & 0.06 & 1.00 & 18000 \\ 
      $\lambda_1$ & 0.86 & 0.05 & 0.77 & 0.96 & 1.01 & 370 \\ 
      $\lambda_2$ & 1.35 & 0.07 & 1.21 & 1.50 & 1.06 & 44 \\ 
      $\nu_1$ & 0.03 & 0.04 & -0.04 & 0.10 & 1.00 & 730 \\ 
      $\nu_2$ & 0.10 & 0.05 & 0.01 & 0.20 & 1.04 & 66 \\ 
      $\sigma_{y_1}^2$ & 0.28 & 0.02 & 0.24 & 0.32 & 1.04 & 65 \\ 
      $\sigma_{y_2}^2$ & 0.34 & 0.03 & 0.29 & 0.40 & 1.04 & 68 \\ 
      $\sigma_{y_3}^2$ & 0.23 & 0.04 & 0.15 & 0.31 & 1.10 & 30 \\ 
       \hline
    \end{tabular}
\end{table}

\subsection{Discussion of Illustration 5}

\begin{figure}[ht!]
    \centering
    \includegraphics[width=\textwidth]{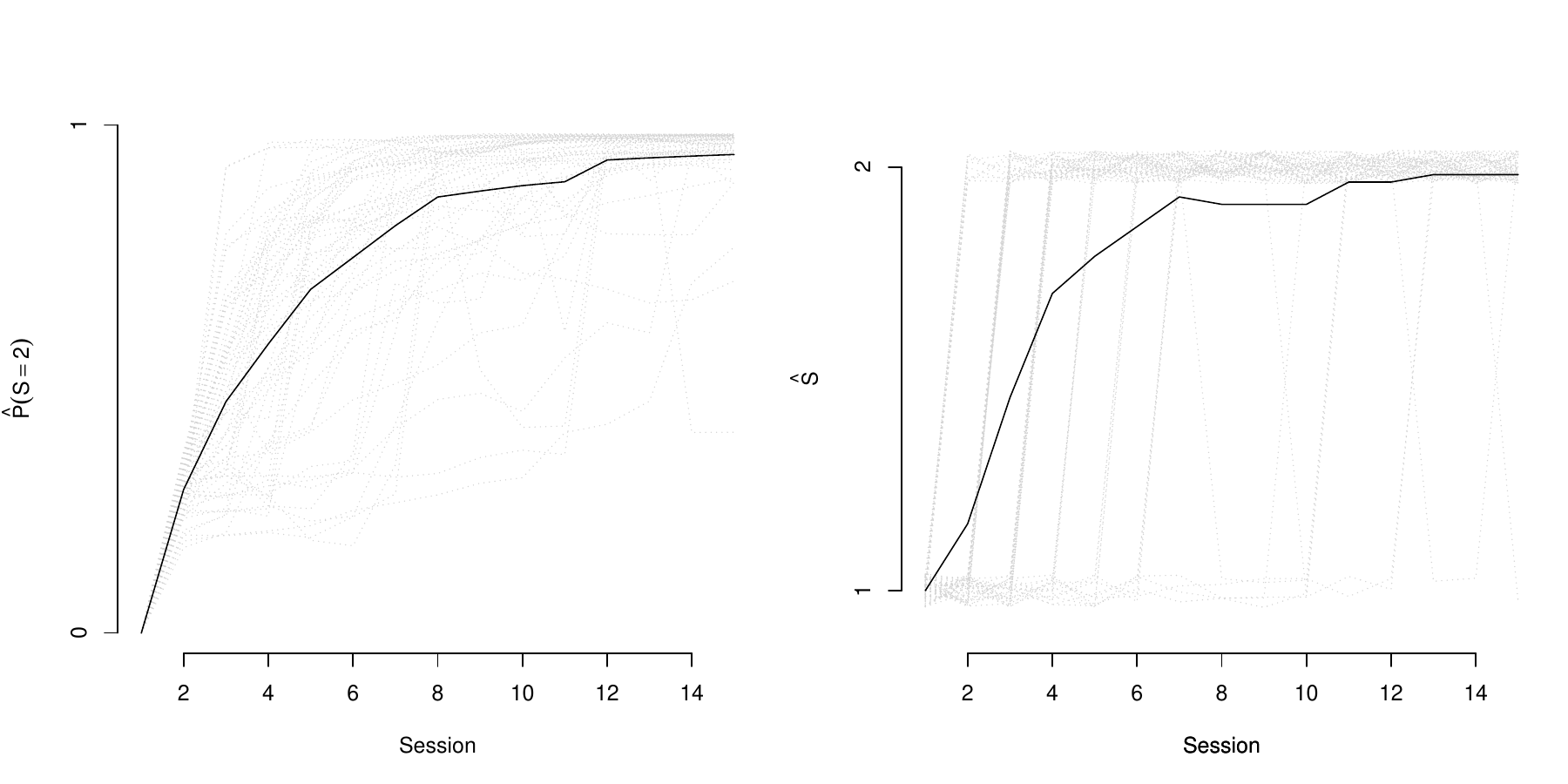}
    \caption{Left: Individual probabilities to transition to state $S_2$ (or remain in $S_2$ after transition) in grey and the averaged probability in black. 
    Right: All patients start in state $S_1$ but move to state $S_2$ over time with more than 75\% of all patients switching after time point 6 (for illustrative purposes, state membership was jittered). 
    }
    \label{fig:dlcsem_prob}
\end{figure}

Figure~\ref{fig:dlcsem_prob} illustrates the results. In the left panel, individual probabilities \chg{of transitioning into state} $S_2$ (and after transition, the probability \chg{of staying} in state $S_2$) \chg{are} depicted. These probabilities increase for each patient over the sessions. In the right panel, the estimated state membership is illustrated across sessions. \chg{Almost} $80\%$ of the patients have switch at time point \chg{5} with virtually all patients switching until the last session ($9\chg{8}\%$).

\begin{figure}[H]
    \centering
    \includegraphics[scale=0.5,page=1]{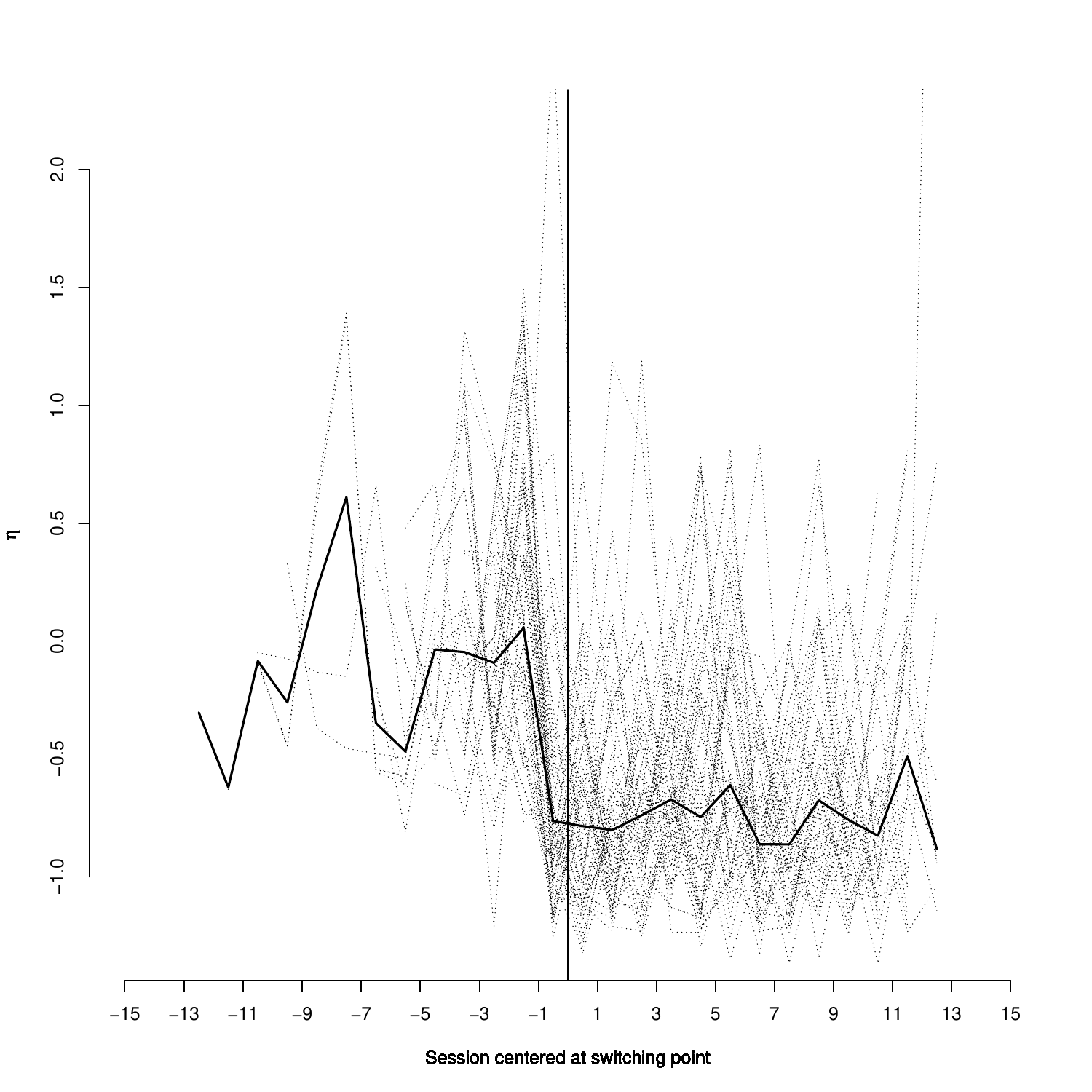}
    \caption{Individual trajectories for the latent anxiety factor based on the DLCSEM for each patient. Trajectories are centered around the estimated individual switching time point. At the switching point, a sharp decline (sudden change) in anxiety is visible for virtually all patients. The bold line indicates the averaged trend in this plot.
    }
    \label{fig:dlcsem_individual}
\end{figure}

\chg{To illustrate how the state switching relates to} the development of the patients' anxiety, the estimated factor scores are depicted in Figure~\ref{fig:dlcsem_individual}, centered around the individual switching points. Patients show a rather stable initial anxiety on a higher level, a sharp decline in the anxiety factor at the switching point, and then follow a rather stable anxiety level for the remaining time. \chg{Thus, state switching corresponds to a sudden improvement in the patients’ mental health outcome, the level of anxiety. These findings are consistent with H4, which posited sudden changes in anxiety levels at specific time points for individual patients.}

\chg{Moreover, the timing of switching can be meaningfully interpreted.} The CBT intervention in this study used the standard MAW protocol \cite{Zinbarg2006}, which typically includes a foundation and skills training in the initial three to four sessions and starts with cognitive restructuring (an effective method to reduce anxiety) in sessions four to five. While the individual onset of this restructuring was not available in the data set, the rather similar time point of switching around time point \chg{5} for most patients \chg{aligns well with the expected timing of this therapeutic component. However, it is important to emphasize that although the data originate from a randomized clinical trial, the present analyses do not incorporate treatment allocation and therefore treat the data as observational. Consequently, the inferred switching represents a model-based description of latent trajectory dynamics rather than a causal effect of specific therapeutic components.}

\subsection{\chg{Learnings Regarding the Procedure}}

\chg{We have now reached the final step of the modeling procedure: formulating the full DLCSEM. At this stage, the model should formalize a specific hypothesis grounded in theoretical considerations and informed by observations from the simpler models.}

\chg{In Illustration 5, the observed decline in patient anxiety over time, together with theoretical expectations about therapy processes, motivated the hypothesis that improvements may accumulate gradually, emerging only after several sessions and subsequently leading to pronounced changes in outcomes. In other applied contexts, analogous hypotheses might involve delayed intervention effects, developmental turning points, or relapse processes.}

\chg{\paragraph{Specific Model Decisions} Even with such a confirmatory approach, modeling decisions in DLCSEM require careful attention. Misspecifications can easily lead to non-convergence or unstable estimates (see the Discussion and Appendix~\ref{app:convergence} for further details). In the following, we highlight some of the main challenges specific to Illustration 5.}

\chg{First, model stability depends on a clear and data-supported definition of the states. In Illustration 5, the two states are distinguished through the censored prior $\delta_\alpha$, which constrains state $S_2$ to represent lower anxiety levels than state $S_1$. Without this censoring, the model becomes unidentified and cannot be estimated.}

\chg{Stability can be further enhanced by specifying a more informative prior for $\delta_\alpha$. Estimation is sensitive to its magnitude: if $\delta_\alpha$ is large enough to clearly separate the states, convergence tends to be stable; if it is small or inconsistent with the data, estimation problems are more likely to occur.}

\chg{At the same time, imposing a particular state structure through an informative prior is a substantive modeling decision that must be theoretically and empirically justified. If the imposed state definition contradicts the observed data, the resulting estimates may be unstable or practically meaningless. For example, applying opposed censoring (i.e., defining state $S_1$ as representing lower anxiety) conflicts with the observed decrease in anxiety and leads to almost no individuals switching states.}

\chg{Second, model complexity must be calibrated relative to sample size. In our example, the number of patients (level-2 units) is the primary limiting factor. To ensure stable estimation, we fixed the covariance between random effects. Including additional covariances did not meaningfully alter the results but increased instability.}

\chg{Finally, cross-state equality constraints are crucial for both the interpretation of the factor structure and the stability of estimation. For example, allowing the residual variance of the within-level factor to differ across states can lead to non-convergence.}

\chg{Potential issues do not arise in every case, but they become more likely when the data provide only weak or ambiguous evidence for differences between states. As a general takeaway, modeling decisions should be guided by observable data patterns while ensuring that the resulting model provides a good fit and remains interpretable.}

%% file: fig/06_dlcsem.tex
\begin{figure}[H]
\begin{center}
\resizebox{.7\textwidth}{!}{
\begin{tikzpicture}[>=stealth,semithick]
	
	\node[ov] (y11) at (0,1)  {$y_{it1}$};
	\node[ov] (y12) at (1,1)  {$y_{it2}$};
	\node[ov] (y13) at (2,1)  {$y_{it3}$};
	\node[lv] (eta1) at (1,0)  {$\eta_{it}$};
	\node[lr] (zeta1) at (0,-1)  {$\zeta_{it}$};

	\node[ov,scale=0.8] (y21) at (3.5,1)  {$y_{i(t+1)1}$};
	\node[ov,scale=0.8] (y22) at (5,1)  {$y_{i(t+1)2}$};
	\node[ov,scale=0.8] (y23) at (6.5,1)  {$y_{i(t+1)3}$};
	\node[lv,scale=0.6] (eta2) at (5,0)  {$\eta_{i(t+1)}$};
	\node[lr,scale=0.8] (zeta2) at (4,-1)  {$\zeta_{i(t+1)}$};
	
	\path[->] 
	(eta1)  edge node[above,scale=0.8] {} (y11)
	(eta1)  edge node[above,scale=0.8] {} (y12)
	(eta1)  edge node[above,scale=0.8] {} (y13)
    (eta2)  edge node[above,scale=0.8] {} (y21)
	(eta2)  edge node[above,scale=0.8] {} (y22)
	(eta2)  edge node[above,scale=0.8] {} (y23)
    ;
	\path[->] 
	(eta1)  edge node[scale=1] {$\bullet$}  (eta2)
	;

    \node[re] (re1a) at (3,-.5)     {$\beta_{is}$};
    
	\path[->] 
	(zeta1)  edge node[above,scale=0.8] {} (eta1)
	(zeta2)  edge node[above,scale=0.8] {} (eta2)
	;

    \node[re] (re1) at (1,-1.5)     {$\alpha_{is}$};
    \node[re] (re2) at (5,-1.5)     {$\alpha_{is}$};
    
    \path[->] 
	(re1)  edge node[scale=1] {$\bullet$} (eta1)
	(re2)  edge node[scale=1] {$\bullet$} (eta2)
	;

	\draw[dashed] (-2,-2) -- (8,-2);
	\node (text1) at (7,-1) {within};
	\node (text1) at (7,-3) {between};
	
	\node[re] (u0) at (2,-3)      {$\alpha_{is}$};
   \node[re] (u1) at (4,-3)      {$\beta_{is}$};


	\end{tikzpicture}
}
\end{center}
\caption{Path diagram for the DLCSEM using AR(1) structure for a single latent factor with $T$ time points and $N$ persons. The two states $s=S_1$ and $s=S_2$ are modeled such that persons who respond to the treatment receive higher scores in the latent factor.}
\label{fig:02_arma11}
\end{figure}

%% file: 07_fusion.tex
\section{Example of Structurally Different State Models}

\subsection{Data and Goal of Illustration 6}

\chg{For this final illustration, we build on the observations from Illustration 1 regarding the WAI's dimensional structure over time. We found that the correlations between the WAI factors were smaller at the first time point compared to the last. In the original paper \cite{Flueckiger2021}, we also observed that the eigenvalue pattern changes over the course of therapy, with a predominant eigenvalue emerging over time. This suggests that the subscales gradually converge, indicating a shift from a multidimensional structure toward a more unidimensional representation, which we refer to as "fusion". While Illustration 1 focused on selected time points, we now include all time points to explore the dynamic shift in the response pattern to test H5.}

\subsection{Implementation of Fusion Model}

The two latent states are now represented by
\numberedtabx{
    $\eta_{S_1it}$  
    &\quad\quad\texttt{eta.S1[i,t,1:3]}\\ 
    \hspace{0.5cm} $\sim\mathcal{MVN}( \boldsymbol{\mu}_{\eta_{S_1it}}, \boldsymbol\Sigma_{\eta_{S_1}}^2 ) $  
    &\quad\quad\texttt{\hspace{1cm}$\sim$ dmnorm(mu.eta.S1[i,t,1:3], }\\ 
    &\hspace{3.9cm}\texttt{psi.eta.S1[1:3,1:3])}\\
    $\eta_{S_2it} \sim \mathcal{N}( \mu_{\eta_{S_2it}}, \sigma_{\eta_{S_2}}^2 ) $  
    &\quad\quad\texttt{eta.S2[i,t] $\sim$ dnorm(mu.eta.S2[i,t],}\\ 
    &\hspace{5.3cm}\texttt{psi.eta.S2)}\\
}{code:dlc_eta_distr_fusion}
where $\eta_{S_1it}$ includes the three WAI subfacets; and $\eta_{S_2it}$ only includes the global WAI factor. Each factor follows an AR(1) structure with a random intercept, implemented analogously to Equation~\Cref{code:dlc_mu_distr}. 
%
%
Factor loadings and residual variances were specified state-specifically.
The probabilities $\texttt{P1}$ were modeled very similarly to \Cref{code:logistic_predictor} and used all three factors from state $S_1$ to predict a switch to state $S_2$. $\texttt{P2}$ was again implemented using a uniform prior as specified in \cref{code:fixed_transition}. \chg{In order to improve stability and convergence, we set the intercepts to zero, which was in line with the data that were centered with the mean of the first measurement occasion (see Data subsection).} 
All other priors were identical to the previous illustrations.

\chg{The path diagram in \Cref{fig:fusion1} displays the factor models for both states: state $S_1$ is specified as a three-factor model, whereas state $S_2$ is specified as a single-factor model (left panel). The right panel depicts the autocorrelations and the HMSM.}

\begin{figure}[H]
    \begin{center}
        \resizebox{!}{.5\textheight}{\input{fig/07_fusion}}
        \resizebox{!}{.5\textheight}{\input{fig/07_fusion2}}
        \caption{Path diagram for the DLCSEM that implements a three-factor model for state $S_1$ and single factor model for state $S_2$. \chg{Left: Multilevel measurement models.} Right: Autoregressive models and HMSM..}
        \label{fig:fusion1}
    \end{center}
\end{figure}

\subsection{Results of Illustration 6}

\chg{Posterior estimates for the shared parameters in the fusion model are presented in \Cref{tab:fusion_shared}. The transition probability from state $S_2$ to state $S_1$ ($P_{21}$) was very low ($0.02$), indicating that individuals were unlikely to switch back once in state $S_2$. Among the transition parameters, $b_{21}$ was substantial ($2.25$), while $b_{22}$, $b_{23}$, and $b_{24}$ were close to zero with wide credible intervals, suggesting limited or uncertain lagged effects. Residual variances for the observed indicators ranged from $0.18$ to $0.52$, reflecting moderate variability after accounting for the factor structure.}

\chg{Parameter estimates for state $S_1$, specified as a three-factor model, are presented in \Cref{tab:fusion_state1}. Standardized factor loadings ranged between $0.46$ and $1.67$. Factor correlations were of medium magnitude, ranging from $0.54$ to $0.69$ 
at the within level and slightly smaller at the between level ($0.29$ to $0.50$)
. The factor variances indicated moderate to high ICCs for all three factors, amounting to $0.50$, $0.55$, and $0.69$, respectively. None of the regression coefficients suggested a substantial lagged effect, as all credible intervals included zero.}

\begin{table}[ht!]
\centering
 \caption{Illustration 6 (part 1): Posterior estimates (posterior mean, standard deviation (SD), and 95\% credible interval defined by the 2.5\% and 97.5\% quantiles), $\hat{R}$ statistic and effective sample size ($n_{eff}$) for the fusion model: transition probability from state $S_2$ in state $S_1$ ($P_{21}$), transition parameters $b_{21}$ and $b_{22}$, and residual variances ($\sigma_{y_j}^2$).}
\label{tab:fusion_shared}
\begin{tabular}{lrrrrrr}
\hline
 & mean & sd & 2.5\% & 97.5\% & $\hat{R}$ & $n_{\text{eff}}$ \\
\hline
$P_{21}$ & 0.02 & 0.01 & 0.00 & 0.04 & 1.02 & 130 \\
$b_{21}$ & 2.25 & 0.26 & 1.77 & 2.81 & 1.03 & 110 \\
$b_{22}$ & -0.24 & 0.76 & -1.75 & 1.24 & 1.01 & 270 \\
$b_{23}$ & -0.68 & 0.86 & -2.35 & 1.01 & 1.01 & 250 \\
$b_{24}$ & -1.07 & 0.74 & -2.55 & 0.34 & 1.01 & 470 \\
$\sigma^2_{y_1}$ & 0.29 & 0.02 & 0.25 & 0.32 & 1.01 & 670 \\
$\sigma^2_{y_2}$ & 0.18 & 0.02 & 0.15 & 0.21 & 1.02 & 150 \\
$\sigma^2_{y_3}$ & 0.24 & 0.02 & 0.21 & 0.27 & 1.00 & 730 \\
$\sigma^2_{y_4}$ & 0.46 & 0.03 & 0.40 & 0.52 & 1.02 & 170 \\
$\sigma^2_{y_5}$ & 0.27 & 0.02 & 0.23 & 0.31 & 1.01 & 460 \\
$\sigma^2_{y_6}$ & 0.22 & 0.02 & 0.19 & 0.26 & 1.01 & 640 \\
$\sigma^2_{y_7}$ & 0.37 & 0.03 & 0.32 & 0.42 & 1.03 & 110 \\
$\sigma^2_{y_8}$ & 0.26 & 0.02 & 0.22 & 0.30 & 1.06 & 49 \\
$\sigma^2_{y_9}$ & 0.52 & 0.03 & 0.46 & 0.58 & 1.00 & 11000 \\
\hline
\end{tabular}
\end{table}

\begin{table}[ht!]
\centering
 \caption{Illustration 6 (part 2): Posterior estimates (posterior mean, standard deviation (SD), and 95\% credible interval defined by the 2.5\% and 97.5\% quantiles), $\hat{R}$ statistic and effective sample size ($n_{eff}$) for state $S_2$ in the fusion model: intercepts $\alpha_{\cdot S_1}$ and autoregressive coefficients $\beta_{\cdot\cdot S_1}$,  covariances of person-specific random intercept ($\boldsymbol{\Sigma}^2_{\zeta_{21},S_1}$), within level factor residual covariances ($\Sigma_{\eta}^2$), and factor loadings ($\lambda_{S_1}$).}
\label{tab:fusion_state1}
\begin{tabular}{lrrrrrr}
\hline
 & mean & sd & 2.5\% & 97.5\% & $\hat{R}$ & $n_{\text{eff}}$ \\
\hline
$\alpha_{1,S_1}$ & 0.00 & 0.00 & 0.00 & 0.00 & 1.00 & 1 \\
$\alpha_{2,S_1}$ & 0.00 & 0.00 & 0.00 & 0.00 & 1.00 & 1 \\
$\alpha_{3,S_1}$ & 0.00 & 0.00 & 0.00 & 0.00 & 1.00 & 1 \\
$\beta_{11,S_1}$ & 0.17 & 0.17 & -0.16 & 0.52 & 1.02 & 110 \\
$\beta_{21,S_1}$ & 0.08 & 0.19 & -0.31 & 0.43 & 1.04 & 64 \\
$\beta_{31,S_1}$ & 0.04 & 0.14 & -0.26 & 0.31 & 1.01 & 530 \\
$\beta_{12,S_1}$ & 0.07 & 0.13 & -0.22 & 0.32 & 1.03 & 210 \\
$\beta_{22,S_1}$ & 0.10 & 0.17 & -0.19 & 0.45 & 1.03 & 130 \\
$\beta_{32,S_1}$ & 0.07 & 0.11 & -0.16 & 0.27 & 1.02 & 170 \\
$\beta_{13,S_1}$ & 0.21 & 0.24 & -0.27 & 0.73 & 1.05 & 63 \\
$\beta_{23,S_1}$ & 0.19 & 0.27 & -0.35 & 0.71 & 1.04 & 79 \\
$\beta_{33,S_1}$ & 0.13 & 0.22 & -0.31 & 0.56 & 1.05 & 82 \\
$\Sigma^2_{\zeta_{21},S_1}[1,1]$ & 0.16 & 0.04 & 0.10 & 0.26 & 1.02 & 140 \\
$\Sigma^2_{\zeta_{21},S_1}[2,1]$ & 0.04 & 0.03 & -0.01 & 0.10 & 1.02 & 120 \\
$\Sigma^2_{\zeta_{21},S_1}[3,1]$ & 0.14 & 0.06 & 0.05 & 0.28 & 1.05 & 63 \\
$\Sigma^2_{\zeta_{21},S_1}[2,2]$ & 0.12 & 0.03 & 0.07 & 0.19 & 1.01 & 220 \\
$\Sigma^2_{\zeta_{21},S_1}[3,2]$ & 0.03 & 0.04 & -0.05 & 0.12 & 1.03 & 98 \\
$\Sigma^2_{\zeta_{21},S_1}[3,3]$ & 0.50 & 0.13 & 0.30 & 0.80 & 1.06 & 52 \\
$\Sigma^2_{\eta,S_1}[1,1]$ & 0.16 & 0.02 & 0.12 & 0.21 & 1.03 & 130 \\
$\Sigma^2_{\eta,S_1}[2,1]$ & 0.08 & 0.01 & 0.05 & 0.11 & 1.01 & 420 \\
$\Sigma^2_{\eta,S_1}[3,1]$ & 0.13 & 0.02 & 0.09 & 0.17 & 1.01 & 810 \\
$\Sigma^2_{\eta,S_1}[2,2]$ & 0.10 & 0.02 & 0.07 & 0.14 & 1.02 & 210 \\
$\Sigma^2_{\eta,S_1}[3,2]$ & 0.08 & 0.02 & 0.05 & 0.11 & 1.01 & 880 \\
$\Sigma^2_{\eta,S_1}[3,3]$ & 0.22 & 0.03 & 0.16 & 0.29 & 1.03 & 92 \\
$\lambda_{1,S_1}$ & 1.44 & 0.09 & 1.26 & 1.63 & 1.02 & 130 \\
$\lambda_{2,S_1}$ & 1.25 & 0.09 & 1.08 & 1.43 & 1.01 & 310 \\
$\lambda_{3,S_1}$ & 1.43 & 0.13 & 1.19 & 1.68 & 1.01 & 210 \\
$\lambda_{4,S_1}$ & 1.67 & 0.15 & 1.38 & 1.98 & 1.02 & 140 \\
$\lambda_{5,S_1}$ & 0.98 & 0.06 & 0.87 & 1.09 & 1.09 & 35 \\
$\lambda_{6,S_1}$ & 0.46 & 0.06 & 0.35 & 0.58 & 1.02 & 130 \\
\hline
\end{tabular}
\end{table}

\chg{Parameter estimates for state $S_2$, specified as a single-factor model, are shown in \Cref{tab:fusion_state2}. Standardized factor loadings were comparable to those observed in state $S_1$, ranging from $0.70$ to $1.29$. The ICC for state $S_2$ was larger ($0.83$), indicating that alliance ratings in this state were more strongly driven by inter-individual differences than by temporal fluctuations. The autocorrelation was high ($\beta_{S_2} = 0.77$), implying that alliance ratings at session $t-1$ were strongly predictive of ratings at session $t$ following a state switch.}

\begin{table}[H]
\centering
 \caption{Illustration 6 (part 3): Posterior estimates (posterior mean, standard deviation (SD), and 95\% credible interval defined by the 2.5\% and 97.5\% quantiles), $\hat{R}$ statistic and effective sample size ($n_{eff}$) for state $S_2$ in the fusion model: intercept ($\alpha_{S_2}$ and autoregressive coefficient ($\beta_{S_2}$), variance of person-specific random intercept ($\boldsymbol{\sigma}^2_{\zeta_21}$), within level factor residual variance ($\sigma_{\eta}^2$), and factor loadings ($\lambda_{S_2}$).}
\label{tab:fusion_state2}
\begin{tabular}{lrrrrrr}
\hline
 & mean & sd & 2.5\% & 97.5\% & $\hat{R}$ & $n_{\text{eff}}$ \\
\hline
$\alpha_{S_2}$ & 0.72 & 0.09 & 0.55 & 0.89 & 1.01 & 360 \\

$\beta_{S_2}$ & 0.77 & 0.10 & 0.56 & 0.94 & 1.00 & 730 \\

$\sigma^2_{\zeta_{21},S_2}$ & 0.19 & 0.06 & 0.11 & 0.33 & 1.01 & 210 \\

$\sigma^2_{\eta,S_2}$ & 0.04 & 0.01 & 0.03 & 0.05 & 1.00 & 1200 \\

$\lambda_{1,S_2}$ & 0.86 & 0.04 & 0.79 & 0.94 & 1.01 & 590 \\
$\lambda_{2,S_2}$ & 0.91 & 0.04 & 0.84 & 0.99 & 1.01 & 540 \\
$\lambda_{3,S_2}$ & 1.29 & 0.06 & 1.18 & 1.41 & 1.01 & 270 \\
$\lambda_{4,S_2}$ & 0.70 & 0.04 & 0.62 & 0.78 & 1.01 & 280 \\
$\lambda_{5,S_2}$ & 0.71 & 0.04 & 0.64 & 0.78 & 1.00 & 2700 \\
$\lambda_{6,S_2}$ & 0.81 & 0.04 & 0.72 & 0.89 & 1.00 & 2800 \\
$\lambda_{7,S_2}$ & 1.03 & 0.04 & 0.95 & 1.12 & 1.00 & 1200 \\
$\lambda_{8,S_2}$ & 1.25 & 0.06 & 1.14 & 1.36 & 1.00 & 4500 \\

\hline
\end{tabular}
\end{table}

\subsection{Discussion of Illustration 6}

\begin{figure}[H]
    \centering
    \includegraphics[width=\textwidth]{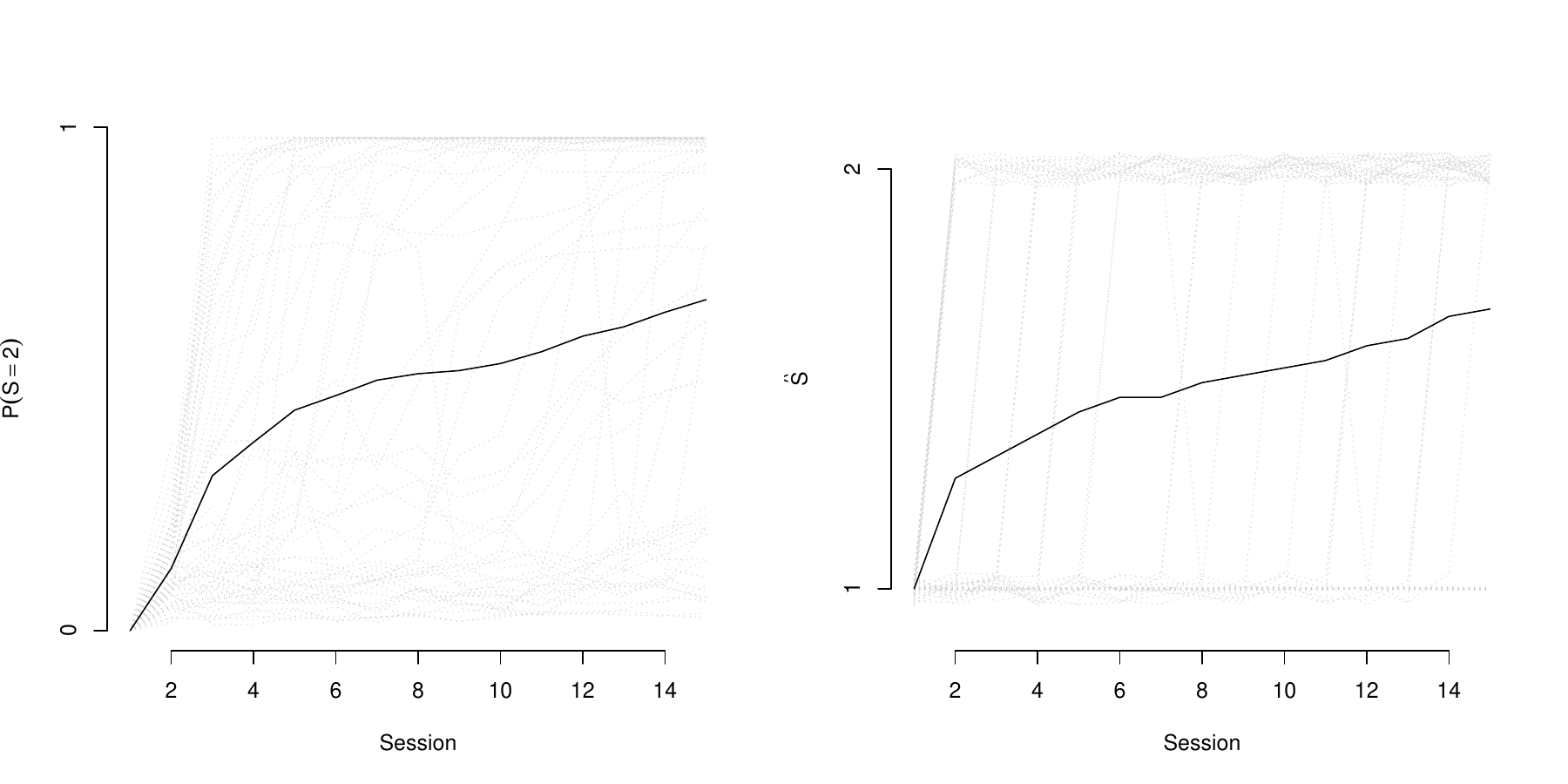}
    \caption{Left: Individual probabilities to transition to state $S_2$ (or remain in $S_2$ after transition) in gray and the averaged probability in black. 
    Right: All patients start in state $S_1$, but about two-thirds move to state $S_2$ over time (for illustrative purposes, state membership was jittered). 
    }
    \label{fig:fusion2}
\end{figure}

\Cref{fig:fusion2} replicates the \Cref{fig:dlcsem_prob} from Illustration 5. It shows a lower and slower switching behavior than in Illustration 5. A total of $66\%$ of the patients actually switch. The average time point of switching is at session $5.2$. The left panel shows that the individual probabilities of those who switch states often increase rapidly at the beginning (session 2). The probabilities of those who remain in state $S_1$ stabilize at a rather low level. The right panel shows that switches back to state $S_1$ are rather unlikely (with an estimated $P_{2}=0.02$). 

If we compare the time points of switching from Illustrations 5 and 6, we only find a correlation of 0.02 between switching to a sudden gain (Illustration 5) and a dimensional shift (Illustration 6). The proposed models and states, hence, need to be interpreted quite differently. \chg{While Illustration 5 targets the development of the health outcome measured by the BAI, Illustration 6 focuses on the evolution of the therapeutic alliance between patient and therapist, measured by the WAI.}

\chg{These results are consistent with previous observations in the CFA and with \citet{Flueckiger2021}. They support H5, indicating a dimensional shift in the response pattern—from a multidimensional factor with three subscales to a unidimensional factor. Although this phenomenon requires further research, the shift may reflect a consolidation of the alliance over time, in which initial variability across alliance dimensions becomes integrated into a coherent, overarching relational pattern.}

\chg{To better understand the mechanisms underlying these dimensional shifts, we examined the logistic regression within the HMSM. The small and uncertain transition parameters ($b_{22}$, $b_{23}$, $b_{24}$) indicate that none of the latent factors significantly influenced whether patients remained in state $S_1$. This suggests that state transitions were driven primarily by the pattern of responses across items, rather than by the absolute scores on the alliance facets. In other words, the configuration or correlations between items determined whether a patient switched states, rather than whether their ratings were generally higher or lower.}

\subsection{\chg{Learnings Regarding the Procedure}}

\chg{Illustration 6 exemplifies how hypotheses about the data structure can be translated into DLCSEM. As discussed in the CFA section, one possible approach is to compare cross-sectional models at selected time points. Evidence for changes in dimensionality can then inform the definition of longitudinal latent states. For example, if the factor structure differs between early and later measurements, well-fitting models from these respective periods can be used to define the corresponding latent states in the DLCSEM.}

\chg{In the final step, the DLCSEM can be used to validate the observed changes in factor structure over time, bringing the analysis back to a confirmatory framework. If the model produces reasonable state switches and interpretable parameter estimates, the hypothesized dimensional shift is supported. While many psychometric tests are validated in cross-sectional settings, there are situations in which the underlying factor structure may evolve. The hypothesis of a dimensional shift can, for example, be extended to contexts such as attention, cognitive performance, or symptom trajectories in clinical psychology.}

\chg{\paragraph{Specific Model Decisions} As discussed for Illustration 5, modeling choices regarding the HMSM are critical, particularly the definition of the states, because these decisions directly influence both interpretability and model convergence. In Illustration 6, the states are clearly defined by differences in the factor structure, which provides a natural basis for distinguishing them. Nevertheless, potential issues may still arise, such as label switching or sensitivity to initial parameter values, especially if the state definitions do not closely correspond to the underlying data structure.}

\chg{A useful strategy in practice is to assess the robustness of the results by running alternative model specifications. For example, one can reverse or modify the definition of the states to check whether the main patterns in the data hold.}

\chg{In an additional illustrative analysis (Appendix~\ref{app:dsem_illu6}), we reversed the state specification, starting patients in the single-factor state and allowing them to switch to the three-factor state. The data patterns quickly overrode this forced structure, producing results very similar to the original model. Across specifications, state recovery was reasonably accurate (about 0.8; see Figures~\ref{fig:defus1} and \ref{fig:defus2}), suggesting that the main patterns may be relatively robust to alternative state definitions in this case.}

%% file: fig/07_fusion.tex
\begin{tikzpicture}[>=stealth,semithick]
	
			
            \node[ov] (y4) at (3,2)  {$y_{it4}$};
			\node[ov] (y5) at (4,2)  {$y_{it5}$};
			\node[ov] (y6) at (5,2)  {$y_{it6}$};
			\node[ov] (y7) at (6,2)  {$y_{it7}$};
			\node[ov] (y8) at (7,2)  {$y_{it8}$};
			\node[ov] (y9) at (8,2)  {$y_{it9}$};
			\node[ov] (y10) at (9,2)  {$y_{it10}$};
			\node[ov] (y11) at (10,2)  {$y_{it11}$};
			\node[ov] (y12) at (11,2)  {$y_{it12}$};
			
			\node[lv] (xi121) at (4,0)      {$\eta_{it1}$};
			\node[lv] (xi131) at (7,0)      {$\eta_{it2}$};
			\node[lv] (xi141) at (10,0)      {$\eta_{it3}$};
			
			\node[lv] (xi221) at (7,4)       {$\eta_{it}$};

            \node[lr] (re11) at (4,-.55)     {$\bullet$};
            \node[lr] (re12) at (7,-.55)     {$\bullet$};
            \node[lr] (re13) at (10,-.55)     {$\bullet$};
            \node[lr] (re2) at (7,4.45)     {$\bullet$};
            
			\path[->] 
			(xi121) edge (y4)
			(xi121) edge (y5)
			(xi121) edge (y6)
			(xi131) edge (y7)
			(xi131) edge (y8)
			(xi131) edge (y9)
			(xi141) edge (y10)
			(xi141) edge (y11)
			(xi141) edge (y12)
            ;
            
			\path[->] 
			(xi221) edge (y4)
			(xi221) edge (y5)
			(xi221) edge (y6)
			(xi221) edge (y7)
			(xi221) edge (y8)
			(xi221) edge (y9)
			(xi221) edge (y10)
			(xi221) edge (y11)
			(xi221) edge(y12)
            ;
			

			\path[<->] (xi121)  edge [bend right=50, looseness=1] (xi131);
			\path[<->] (xi121)  edge [bend right=50, looseness=1] (xi141);
			\path[<->] (xi131)  edge [bend right=50, looseness=1] (xi141);

      
			\node[lv] (xi122) at (4,-6)      {$\alpha_{i1}$};
			\node[lv] (xi132) at (7,-6)      {$\alpha_{i2}$};
			\node[lv] (xi142) at (10,-6)      {$\alpha_{i3}$};
			
			\node[lv] (xi222) at (7,-4)       {$\alpha_{i}$};

			\path[<->] (xi122)  edge [bend right=50, looseness=1] (xi132);
			\path[<->] (xi122)  edge [bend right=50, looseness=1] (xi142);
			\path[<->] (xi132)  edge [bend right=50, looseness=1] (xi142);


        \node (a00) at (3,4)  {\emph{within}};
		\node (a00) at (3,-4)  {\emph{between}};
		\draw[dotted] (0,2) -- (12,2);
		\node (St0) at (1,2.5)  {$S_{it}=2$};
		\node (St1) at (1,1.5)  {$S_{it}=1$};
        \draw[dotted] (0,-5) -- (12,-5);
	      \node (St0) at (1,-4.5)  {$S_{it}=2$};
	    \node (St1) at (1,-5.5)  {$S_{it}=1$};
			
      \draw[dashed] (0,-8.5) rectangle (12,8);

	\end{tikzpicture}

%% file: fig/07_fusion2.tex
\begin{tikzpicture}[>=stealth,semithick]
			
			
			\node (St1) at (2,7.5)  {$S_{t}=1$};
			\node (St0) at (2,.5)  {$S_{t}=2$};
			\node (St1) at (2,-2.5)  {HMSM};
			\draw[dotted] (-2,1) -- (6,1);
			\draw[dotted] (-2,-2) -- (6,-2);
			\draw[dashed] (-2,-8.5) rectangle (6,8);

            \node[lv] (xi111) at (0,6)      {$\eta_{1it1}$};
			\node[lv] (xi211) at (0,4)      {$\eta_{1it2}$};
			\node[lv] (xi311) at (0,2)      {$\eta_{1it3}$};
			
			\node[lv,scale=0.8] (xi121) at (4,6)      {$\eta_{1i(t+1)1}$};
			\node[lv,scale=0.8] (xi221) at (4,4)      {$\eta_{1i(t+1)2}$};
			\node[lv,scale=0.8] (xi321) at (4,2)      {$\eta_{1i(t+1)3}$};

			\path[->] 
			(xi111)  edge (xi121)
			(xi211)  edge (xi121)
			(xi311)  edge (xi121)
			(xi111)  edge (xi221)
			(xi211)  edge (xi221)
			(xi311)  edge (xi221)
			(xi111)  edge (xi321)
			(xi211)  edge (xi321)
			(xi311)  edge (xi321)
			;
      \path[<->,fill=lightgray,draw=lightgray] 
      (xi111)  edge [bend right=50, looseness=1] (xi211)
      (xi111)  edge [bend right=50, looseness=1] (xi311)
      (xi211)  edge [bend right=50, looseness=1] (xi311)
      ;
            
            \node[lv] (xi112) at (0,-1)      {$\eta_{1it}$};
            \node[lv,scale=0.8] (xi122) at (4,-1)      {$\eta_{1i(t+1)}$};

			\path[->] 
			(xi112)  edge (xi122)
			;

            \node[lv] (xi111) at (0,-4)      {$\eta_{1it1}$};
			\node[lv] (xi211) at (2,-4)      {$\eta_{1it2}$};
			\node[lv] (xi311) at (4,-4)      {$\eta_{1it3}$};
			
			\node[lv,scale=0.8] (S1) at (2,-7)  {$S_{i(t+1)}=1$};
      
            \path[->] 
			(xi111)  edge (S1)
			(xi211)  edge (S1)
			(xi311)  edge (S1)
			;
\end{tikzpicture}

%% file: 08_discussion.tex
\section{Discussion}

As ILD become increasingly standard in research and applied fields, the corresponding methodological advancements need to evolve accordingly. DLCSEM provides extensive capabilities for capturing and analyzing complex \chg{temporal processes, making it highly suitable for the analytical demands of ILD.} Our goal in this tutorial was to illustrate the potential applications of DLCSEM as a method that surpasses traditional approaches in flexibility and analytical power, offering a comprehensive yet accessible introduction to the topic.

We demonstrated how to construct sophisticated latent variable models by starting with foundational techniques, \chg{namely} Factor Analysis, \chg{Multilevel Modeling,} Time Series Analysis, \chg{and Regime Switching,} progressively building the necessary framework to arrive at DSEM and DLCSEM. \chg{These steps were accompanied by real-data illustrations, with full code available in the accompanying GitHub repository. We presented guidelines detailing how these preliminary analyses can serve to formulate hypotheses for DLCSEM and to derive a final model capable of testing them.}

\subsection{\chg{Model Complexity as a Challenge}}

\chg{While we aim to encourage researchers to employ DSEM and DLCSEM, we also wish to highlight important challenges associated with these models. Many of the challenges related to DSEM and DLCSEM arise directly from their key advantage: flexibility. The large number of modeling choices can easily lead to highly complex models with many parameters.}

\chg{This increases the demands on model identification, estimation, and interpretation. In this context, it is useful to distinguish structural identifiability, where parameters are intrinsically non-identifiable due to the model’s structure, from practical identifiability, where the data are not informative enough to estimate parameters with adequate precision \cite{WIELAND202160}. The former requires, for example, standard identification constraints in the factor model and a sufficient number of measurement occasions to identify the time structure, depending on the number of lags. The latter directly relates to estimation difficulties, which we discuss in the following.} 

\subsection{\chg{Estimation Difficulties}}

\chg{The combination of multilevel structure, temporal dynamics, and latent state switching results in complex posterior distributions that can be difficult to sample efficiently, increasing vulnerability to convergence difficulties in MCMC estimation. In Appendix~\ref{app:convergence}, we provide a detailed list of common sources of nonconvergence, along with suggestions on how to address them, building and extending on the strategies described in \citet{Asparouhov2022}. In addition, we provide some insights into special characteristics of JAGS that can also cause problems under certain circumstances.}

\chg{In general, it is important to ensure that chains run sufficiently long and mix well, and to perform thorough convergence diagnostics to verify that parameter estimates are reliable. This includes checking effective sample sizes, $\hat{R}$ statistics, and the posterior variability of parameters. Apart from careful model specification, sample size and appropriate prior choices are key factors for stable estimation.}

\chg{\paragraph{Sample Size} 
The DLCSEM jointly estimates continuous latent dynamics, discrete state indicators, and a transition model linking the two, placing substantial demands on sample size at both the between-person and within-person levels.
Simulation evidence for models within this framework \cite{Andriamiarana2023} indicates that a minimum of $N \geq 50$ individuals and $T \geq 25$ measurement occasions is required, with $N > T$ configurations generally preferred. 
The number of individuals primarily drives estimation quality for between-level and transition parameters, whereas $T$ contributes more to the recovery of within-level dynamics and state-dependent intercepts.
These thresholds should be understood as lower bounds: specifications in which the within-level latent process itself drives state switching, or in which factor structures differ across states, introduce additional identification demands likely exceeding the simulated conditions.
Furthermore, reliable estimation of state-specific parameters requires sufficient person-time observations within each latent state; when a state is rarely occupied, individual-level convergence diagnostics should be carefully inspected rather than relying solely on population-level parameter summaries.}

\chg{\paragraph{Priors}Prior specification is a fundamental aspect of Bayesian modeling. In this tutorial, we employed weakly informative priors for practical purposes and did not systematically evaluate alternative choices. Given the modest sample size ($N = 57$ patients), prior assumptions may meaningfully affect the results, although their role in regularization and computational stability remains important even in larger samples. Simulation studies within this framework indicate that (a) minimal sample sizes should exceed $N = 50$ and that (b) while shrinkage priors may improve estimation in more complex models \citep[e.g., for testing measurement invariance with many predictors][]{Brandt2025}, they must be incorporated deliberately and with careful consideration of their regularization properties \cite{Andriamiarana2023, Andriamiarana2025}.}

\subsection{\chg{Interpretation}}

\chg{Beyond statistical considerations, the interpretability and theoretical meaningfulness of latent states must be prudently evaluated. Regime-switching models may identify states that differ statistically but do not correspond to substantively meaningful or theoretically interpretable regimes. Researchers should therefore ensure that identified states align with theoretical expectations or provide new insights that can be justified based on domain knowledge, rather than relying solely on statistical fit.}

\chg{Furthermore, particular caution is required when drawing causal conclusions. The dynamic and switching parameters describe temporal associations and conditional dependencies, but they do not by themselves establish causal relationships. Causal interpretation requires additional assumptions, and without experimental manipulation of the transition mechanism or strong external identification strategies, causal claims based on the transition model remain fundamentally underdetermined. In the language of \citet{Pearl_2009} causal hierarchy, the DLCSEM operates at the first rung, association, and cannot, without further assumptions, ascend to the levels of intervention or counterfactual reasoning that causal claims require.}

\subsection{\chg{Assumptions}}

\chg{As stated in the introduction, we assume equal time intervals between observations. This should be treated with care, as differences in time between measurement periods can introduce bias in the dynamic estimates. In the literature, two possible solutions are typically cited: First, the use of continuous time models directly accounts for unequal intervals \cite{Voelkle2013}. Second, using a simple procedure that inserts missing data for time points that then balances the intervals has been suggested in combination with DSEM and/or DLCSEM \cite{Asparouhov2018}.}

\chg{Our analysis also implicitly made a missing-at-random (MAR) assumption. In JAGS,\footnote{\chg{In JAGS, missing entries in a partially observed node array are supplied as \texttt{NA} and treated as unobserved stochastic nodes; the software then generates samples for these missing values from their posterior predictive distribution conditional on the observed data \cite[see JAGS User Manual, Sec.~3.2]{Plummer2017JAGS}.}} missing values in the observed indicators are handled automatically through data augmentation: any unobserved $y_{itj}$ is treated as an additional unknown quantity and sampled from its full conditional distribution given the current model parameters, latent factor scores, and state memberships. This procedure corresponds to a fully Bayesian data augmentation approach that yields valid inference under the MAR assumption \cite{Gelman2013}.}

\chg{Furthermore, we imposed homoskedasticity and measurement invariance across time, states, and individuals (with the exception of Illustration 6, which explicitly allows switching between latent states that differ in measurement properties and residual variability). Allowing time- or state-specific variances would substantially increase model complexity and may lead to unstable estimation or identification problems (see discussion points above). Such extensions should therefore only be considered when there are strong substantive reasons to expect changing variability (e.g., learning effects or shifts in response consistency).}

\chg{Measurement invariance is essential to ensure that the latent variables retain the same meaning across individuals, time, and states, allowing for consistent interpretation of the latent constructs. At the same time, potential misspecifications such as time-varying cross-loadings or shifts in factor loadings should be examined carefully. In this tutorial, we demonstrated informal diagnostic strategies; more formal approaches exist \citep[e.g.,][]{Mcneish2017}, but methodological tools for systematically extending measurement invariance testing to D(LC)SEM frameworks remain limited and warrant further development.}

\subsection{\chg{Future Research}}

\chg{The computational cost of DLCSEM estimation scales approximately linearly 
in $N$ and $T$, but the number of iterations required for convergence 
can grow substantially as the total number of discrete state indicators and latent trajectories increases. In complex models with multiple latent factors and states, Bayesian MCMC estimation may become prohibitively slow, as existing software typically supports parallelization across chains but offers limited opportunities for within-chain acceleration. These constraints motivate the development of more scalable estimation approaches and potential alternatives to fully Bayesian MCMC for large-scale settings.}

\chg{Model fit assessment remains an open challenge in DLCSEM. Current recommendations largely rely on comparing competing models using information criteria \citep[e.g.,][]{Hamaker2021}. This requires researchers to actively specify and estimate multiple theoretically motivated models. Given the complexity of DLCSEM, this approach may have limited sensitivity for detecting certain types of misspecification, such as violations of measurement invariance or incorrect dynamic structures. Further methodological developments are needed to provide more direct and informative tools for model evaluation.}

%% file: app02_notation.tex
\newpage
\section{Technical background of DSEM and DCLSEM}

In this appendix, we provide the underlying model equations for DSEM and DLCSEM. In addition, we illustrate how the simplified model for illustration 4 is derived from this general framework.

\subsection{DSEM}
\label{App:dsem}

The general mathematical formulation of DSEM is specified for each component of the outcomes $Y_{it}$ and the latent factors $\eta_{it}$, where both of them may be vectors containing several variables \citep[see][]{Asparouhov2018}, .

The equations of the within-level components contain the time structure:
\begin{equation}
\label{eq:dsem_within_level_general}
    \begin{aligned}
        Y_{1it}= & 
        \underbrace{v_{1}}_{\text{Intercept}} + \underbrace{\sum_{l=0}^L \Lambda_{1lit} \eta_{1i(t-l)}}_{\text{factor model}}
        + \underbrace{\sum_{l=1}^L R_{lit} Y_{1i(t-l)}}_{\text{observed variable time structure}}
        + \underbrace{\sum_{l=0}^L K_{1lit} X_{1i(t-l)}}_{\text{covariate impact}}
        + \varepsilon_{1it} \\
        \eta_{1it}= & \underbrace{\alpha_{1}}_{\text{Intercept}}
        + \underbrace{\sum_{l=1}^L B_{1lit} \eta_{1i(t-l)}}_{\text{latent variable time structure}}
        + \underbrace{\sum_{l=0}^L Q_{lit} Y_{1i(t-l)}}_{\text{observed variable time structure}}
        + \underbrace{\sum_{l=0}^L \Gamma_{1lit} X_{1i(t-l)}}_{\text{covariate impact}}
        + \zeta_{1it}.
    \end{aligned}
\end{equation}
Both the within-level outcomes $Y_{1it}$ and the within-level latent factors $\eta_{1it}$ may linearly depend on the current and past within-level outcomes $Y_{1i(t-l)}$, latent factors $\eta_{1i(t-l)}$, and covariates $X_{1i(t-l)}$.\footnote{Extensions to nonlinear models are described in \citet{Kelava2019}.} The number of lags $L$ determines how many past terms are included. 

Random intercepts are not directly included in the above equations but are modeled at the between-level:
\begin{equation}
\label{eq:between-level}
    \begin{aligned}
        & Y_{2i} = \nu_2 + \Lambda_2 \eta_{2i} + K_2 X_{2i} +\varepsilon_{2i} \\
        & \eta_{2i}=\alpha_2+B_2 \eta_{2i}+\Gamma_2 X_{2i}+\zeta_{2i} \\
        & Y_{3t}=\nu_3+\Lambda_3 \eta_{3t}+K_3 X_{3t}+\varepsilon_{3t} \\
        & \eta_{3t}=\alpha_3+B_3 \eta_{3t}+\Gamma_3 X_{3t}+\zeta_{3t} 
    \end{aligned}
\end{equation}
where the random intercept of the items are included in $\varepsilon_{2i}$ and $\varepsilon_{3t}$, for  person-specific and time-specific random effects, respectively. Similarly, random intercepts of the latent factors are included via $\zeta_{2i}$ and $\zeta_{3t}$. 
Note that only one of the fixed intercepts $\nu_1,\nu_2,\nu_3$ for each item is identified and the remaining parameters need to be set to zero (this holds of course for $\alpha_1,\alpha_2,\alpha_3$ analogously).

The Equations~\eqref{eq:between-level} state that each between-level model component -- the person-specific outcome $Y_{2i}$, the person-specific latent factor $\eta_{2i}$, the time-specific outcome $Y_{3t}$, and the time-specific latent factor $\eta_{3t}$ -- may linearly depend on an intercept term ($\nu_2/\alpha_2/\nu_3/\alpha_3$), the latent factors of the same level ($\eta_{2i}/\eta_{3t}$) and some additional covariates of the same level ($X_{2i}/X_{3t}$). The matrices $\Lambda_2$, $B_2, \Lambda_3$, $B_3$, $K_2$, $\Gamma_2$, $K_3$, and $\Gamma_3$ contain the coefficients that need to be estimated, where the $\Lambda$ matrices contain factor loadings of the multi-level measurement model. $\varepsilon_{2i}$, $\zeta_{2i}$, $\varepsilon_{3t}$, and $\zeta_{3t}$ refer to the zero mean residuals. 

\subsection{\chg{Specific Mathematical Model Formulation for Illustration 4}}
\label{app:dsem_illu4}

\chg{Here, we present the mathematical formulation of the simplified DSEM model (\Crefrange{eq:dsem_within_level_general}{eq:between-level}) used in Illustration 4.}

As we do not consider any covariates, all terms containing $X$ may be dropped.
The within-level in \Cref{eq:within-level} becomes\footnote{Here we center $\eta_{1i(t-1)}$ around the intercept $\alpha_{1}+\zeta_{2i1}$ to correct for Nickell's bias \cite{Nickell1981}
    .} 
\begin{equation}
\label{eq:within-level}
    \begin{aligned}
        Y_{1it}= & v_1 + \Lambda_{10} \eta_{1it}+\varepsilon_{1it}\\
        \eta_{1it} = &\alpha_{1} + B_{11i} \eta_{1i(t-1)}^{\text{center}} + \zeta_{1it}.
    \end{aligned}
\end{equation}
All dependencies on previous outcomes are removed from the within-level outcome $Y_{1it}$, leaving only the factor model relating the observations to the latent factor $\eta_{1it}$ at the same time point. 
The factor loadings $\Lambda_{10}$ are assumed to be constant across patient s and time points (i.e. measurement invariance), thus the indices $i$ and $t$ are dropped.\footnote{The constraints for identification introduced 
in the first section apply.} 
The within-level latent factor $\eta_{1it}$ follows an $\operatorname{AR}(1)$ process, capturing the temporal dependency between time $t$ the previous time step $t-1$. Since only a single lag for a single factor is considered, the matrix $B_{1lit}$ reduces to a scalar coefficient $B_{11i}$ which was denoted $\beta$ when the $\operatorname{AR}(1)$ process was introduced in the time series section.

The between-level in \Cref{eq:between-level} reduces to the person-specific random effects. We include to random effects: The latent variables on the between level include the random intercept $\eta_{2i1}$, and the random slope $B_{11i}=\eta_{2i2}$, which we model explicitly as:
\begin{equation}
    \begin{aligned}
        Y_{2i}= & \Lambda_{10} \eta_{2i1}\\
        \eta_{2i1} =& \zeta_{2i1},\\
        \eta_{2i2} =& \alpha_{22} + \zeta_{2i2},
    \end{aligned}
\end{equation}
where we assume cross-level measurement invariance ($\Lambda_2=\Lambda_{10}$). $\zeta_{2i1}$ represents the person-specific component of the latent factor (where the fixed intercept is $\alpha_1$ because $\alpha_{21}=0$).
In this notation $\alpha_{22}$ is the fixed slope parameter in $B_{11i}$ and $\zeta_{2i2}$ is the person-specific random slope. For the sake of clarity and consistency with the above sections (and all figures), we will define $\alpha_{22}=:\beta$ and

\begin{equation}
    \begin{aligned}
        \alpha_{i} :=& \alpha_1 + \zeta_{2i1},\\
        \beta_{i} :=& \beta + \zeta_{2i2},
    \end{aligned}
\end{equation}

Together, the model looks straightforward as 
\begin{equation}
    \begin{aligned}
        Y_{it}= & Y_{1it} + Y_{2i}\\
        =&v_1 + \Lambda_{10}\underbrace{(\eta_{1it}+\zeta_{2i1})}_{\eta_{it}}+\varepsilon_{1it}\\
        \eta_{it} = &\eta_{1it}+\eta_{2i}\\
        &\alpha_{1} + (\beta + \zeta_{2i2}) \eta_{1i(t-1)}^{\text{center}} + \zeta_{1it} + \zeta_{2i1}=\alpha_{i} + \beta_i\eta_{1i(t-1)}^{\text{center}} + \zeta_{1it}
    \end{aligned}
\end{equation}
as expected as an extension of Equation~\ref{eq:centered_ar1}.

\subsection{Extension to DLCSEM}
\label{app:dlcsem}

A general form of the within-level DLCSEM equations \chg{\citep[see][]{Asparouhov2017}} can be expressed as follows:
\begin{equation}
\label{eq:general_dlcsem}
    \begin{aligned}
       {\left[Y_{1 i t} \mid S_{i t}=s\right]} &= v_{1s} + \sum_{l=0}^{L}\Lambda_{1ilts}\,\eta_{1i(t-l)} + \sum_{l=0}^{L}R_{ilts}\,Y_{1i(t-l)} + \sum_{l=0}^{L}K_{1ilts}\,X_{1i(t-l)} + \epsilon_{1it}\\[6pt]
        {\left[\eta_{1 i t} \mid S_{i t}=s\right]}  &= \alpha_{1s} + \sum_{l=1}^L B_{1ilts} \eta_{1i(t-l)}
        \sum_{l=0}^{L}Q_{ilts}\,Y_{1i(t-l)} + \sum_{l=0}^{L}\Gamma_{1ilts}\,X_{1i(t-l)} + \zeta_{1it},
    \end{aligned}
\end{equation}
where the additional subscript $s$ denotes the categorical latent variable representing the state for individual $i$ at time point $t$.

The latent variable model can be extended further \chg{\citep[see][]{Kelava2019}} by including an additional term $h_{1 l l^{\prime}}\left(\eta_{1 i(t-l)}, \eta_{1 i(t-l^{\prime})}\right)$, vector of potentially nonlinear functions of latent variables at different lags (i.e., interactions and quadratic effects) and/or semiparametric functions, such as spline, \cite{Perperoglou2019} useful in conditions where the specific functional form is unknown or not appropriate to define in advance \cite{Marcoulides2018}.

%% file: app03_convergence.tex
\newpage

\section{\chg{Model convergence}}
\label{app:convergence}

\chg{In Bayesian estimation, convergence typically refers to the behavior of the MCMC algorithm used to approximate the posterior distribution. A model is considered to have converged when the draws of the MCMC sampler provide an accurate empirical representation of the target posterior: informally, the chains have reached their stationary distribution and are exploring it efficiently \cite{McElreath2020}. In practice, this is assessed using multiple diagnostics \citep[e.g., $\hat{R}$, effective sample size, autocorrelation and trace plots, or checks for divergent transitions;][]{Gelman2013}, and we require that the MC error is small relative to the posterior uncertainty in the parameters of interest.}

\chg{DSEM and DLCSEM are characterized by complex dependence structures, including strong temporal correlations, hierarchical parameter coupling, and (in DLCSEM) latent Markov state dependence. These features can induce
highly correlated, multi-modal, and ``funnel-shaped'' posterior geometries, which substantially slow MCMC mixing and complicate convergence \cite{papaspiliopoulos2007general,Betancourt2017,Gelman2006}. Problems can arise at several levels: weak or partial identifiability of factor loadings and structural parameters \cite{Lee2007}; strong posterior correlations across time points or between levels of the hierarchy \cite{papaspiliopoulos2007general}; label-switching and multimodality in the latent classes \cite{Scott01032002}; and poorly scaled or overly diffuse priors that exacerbate funnel-shaped posteriors \cite{papaspiliopoulos2007general,Gelman2006}. In such settings, careful model specification and parameterization, informed prior choices, and convergence diagnostics are essential to ensure that the MCMC output can be interpreted as an approximation of the underlying posterior distribution.
Many problems and possible approaches to solve these convergence issues in DSEM can be found in \citet{Asparouhov2022}. In this section we want to add to this topic for the DLCSEM. We want to distinguish two sources of non-convergence: model formulation and sampling specific issues.}

\subsection{\chg{Common sources of nonconvergence due to model (mis)-specification}}

\chg{For DLCSEM, particularly poorly specified HMSMs can impact the convergence of the model. This impact often is not only related to the model itself but its relation to sample size and the number of persons assigned to the states.}

\begin{itemize}
  \item \chg{\textbf{Design sample size with DLCSEM requirements in mind.} 
  Simulation results \cite{Andriamiarana2023} show that very small samples lead to biased and unstable estimates, especially for the between-level and the HMSM part. A pragmatic lower bound is about $N = 50$ and $T = 25$, with a clear preference for designs where $N > T$ (e.g., $N = 75$, $T = 25$ rather than $N = 50$, $T = 50$), in order to obtain convergent, efficient estimates for both structural and measurement parameters.}

  \item \chg{\textbf{Treat the HMSM / transition model as a convergence bottleneck.} 
  The latent Markov process is often the most fragile part of DLCSEM: Sensitivity and specificity of class assignment degrade quickly when $N$ and $T$ are too small and when priors on the transition structure are poorly chosen. For the HMSM part, combinations such as $T \ge 25$ and $N \ge 25$ can be acceptable, but robust estimation of both continuous dynamics and discrete regimes is only consistently obtained for designs around $N \ge 50$, $T \ge 25$.}

  \item \chg{\textbf{Consider expected state size.} For models to provide reasonable estimates and convergence, the number of persons in each state needs to be sufficiently large (there are no simulation studies that provide a more in-depth answer on how large). If persons do not switch or switch only very late during the sequence, parameter estimates in this state are typically poor. Model adaptation is necessary, e.g., by constraining the model more strongly to reflect relevant subgroups (i.e. a minimal effect size like in illustration 5). If this is unsuccessful, researchers might need to concede that their model cannot be found in the data pattern.}
  
  \item \chg{\textbf{Avoid diffuse priors for between-level variances and factor loadings.} 
  Diffuse priors on the between-level covariance matrix and factor loadings (e.g., vague inverse-Wishart priors) induce substantial bias and loss of power in small to moderate samples, particularly when between-level indicators have low reliability or factors are highly correlated. Informative or at least thoughtfully chosen weakly informative priors on factor loadings and (co)variances markedly improve estimation accuracy and reduce convergence problems; inaccurate but reasonably informative priors are typically less harmful than diffuse ones \cite{Andriamiarana2023}.}

  \item \chg{\textbf{Use regularizing priors for the transition model, but prefer ridge-type priors.} 
  In multilevel dynamic latent variable models where latent class transitions are governed by logistic regression on (many) predictors, Bayesian regularization is essentially mandatory to avoid overfitting and non-identifiability. Simulation studies \cite{Andriamiarana2025} comparing Bayesian Lasso, adaptive spike-and-slab Lasso, regularized horseshoe, and ridge priors show that ridge priors with moderate shrinkage provide the most stable convergence, accurate posterior means/intervals, and a good balance between power and type I error for the transition coefficients. In contrast, Lasso-type priors tend to overshrink relevant effects, and heavy-tailed priors (Cauchy-like horseshoe variants) can behave poorly in logistic models, especially under separation \cite{Andriamiarana2025}.}

  \item \chg{\textbf{Regularize only where needed and control tail behavior.} 
  For the Markov-switching (logistic) part, priors with light or Student-$t$ tails (e.g., ridge, or carefully tuned regularized horseshoe) are preferable to very heavy-tailed alternatives. These choices improve MCMC stability and avoid infinite or highly unstable posterior moments in the presence of quasi-separation, while still allowing non-zero effects to escape excessive shrinkage. Overly aggressive shrinkage or very heavy tails both exacerbate convergence and mixing issues in the transition model.}

  \item \chg{\textbf{Exploit reparameterizations of the Markov-switching component.} 
  For Stan-type HMC estimators, good convergence in DLCSEM is facilitated by marginalizing over the discrete states and using a Hamilton (forward) filter representation of the Markov chain, so that only continuous parameters are sampled. This reparameterization improves geometry for NUTS, yields better effective sample sizes for the transition coefficients, and avoids pathologies associated with sampling discrete regime indicators directly.}\footnote{\chg{\url{https://mc-stan.org/learn-stan/case-studies/hmm-example.html}}, \chg{\url{ https://mc-stan.org/docs/functions-reference/hidden_markov_models.html}}} 

  \item \chg{\textbf{Combine convergence diagnostics with performance metrics on regimes.} 
  Beyond standard MCMC diagnostics ($\hat{R}$, effective sample size, divergences), simulation studies \cite{Andriamiarana2023} highlight the value of monitoring sensitivity and specificity of latent state recovery and the stability of regime-specific parameter estimates as additional “convergence signals.” Poor classification (low sensitivity/specificity) or strongly unstable regime-specific intercepts and slopes often co-occur with problematic prior choices or underpowered designs, even when formal chain diagnostics look acceptable.}
  
\end{itemize}

\subsection{\chg{Common sources of non-convergence due to JAGS}}

\chg{There are multiple available options for Bayesian modeling, each of them with advantages and disadvantages \cite{Vstrumbelj2024}. We want to highlight some features of JAGS that could be of interest to readers coming from other languages, software, or approaching Bayesian modeling for the first time. In JAGS, each relation denotes a node in the model in terms of other nodes. Taken together, the nodes in the model form a directed acyclic graph.}

\chg{During model compilation, JAGS constructs an internal graph representation of the model in memory. Because the model is defined through this graph structure, the order in which statements are written is irrelevant: rearranging lines that specify nodes does not change the resulting model. This property is uncommon in many widely used programming languages, such as R, Python, and Stan, which follow an imperative paradigm where execution depends on the sequential order of commands. In contrast, JAGS \citep[along with BUGS;][]{Lunn2009} employs a declarative programming paradigm, in which the model is specified by describing the relationships among variables rather than the control flow of computations. For a more detailed discussion of declarative versus imperative programming, see \citet{Leslie2001}.}

\chg{A common situation in which alternative Bayesian software may be preferred arises when models become highly complex or when computational scalability and speed are critical. For example, a comparison between JAGS and Stan in the context of dynamic structural equation modeling is presented in \citet{Hecht2021}. When implementing models in different software environments, it is important to account for differences in parameterization and sampling algorithms, as these can affect both convergence and computational efficiency. In particular, careful attention should be paid to parameter ordering and model specification to ensure proper MCMC execution; see \citet{Asparouhov2010} for a detailed discussion of these issues.}

\begin{itemize}
  \item \chg{\textbf{Identical automatic initial values across chains.}
 When initial values are not supplied, JAGS generates default starting values based on “typical” values implied by the prior distributions. In practice, these defaults may be identical or nearly identical across chains, particularly when models are scripted with the same random number generator settings. In high-dimensional dynamic SEMs, such homogeneous initialization can cause chains to explore the same posterior region, potentially masking multimodality and giving a misleading impression of convergence. Providing chain-specific, overdispersed initial values is therefore strongly recommended.}
  
  \item \chg{\textbf{Insufficient adaptation and burn-in.}
  JAGS relies on an adaptive phase for its samplers; if adaptation is too short or stopped prematurely, samplers may never reach their optimal configuration, leading to slow mixing or apparent non-convergence even for well-specified models. Complex dynamic latent models typically require substantially longer adaptation and burn-in than the defaults.}

  \item \chg{\textbf{Slice-sampler pathologies at the boundaries.}
  The default slice samplers can get “stuck” when priors imply infinite density at the boundary of the support (e.g., \texttt{dgamma} or \texttt{dbeta} with shape parameters $<1$). In practice this can produce errors such as the slicer being stuck at a value with infinite density and effectively halt the exploration of the posterior, unless one truncates or reparameterizes these priors.}

  \item \chg{\textbf{Suboptimal automatic sampler choice.}
  Sampler factories in JAGS assign generic samplers automatically; for complex blocks (e.g., high-dimensional latent states, covariance structures, or logistic components) the chosen sampler may be technically correct but extremely inefficient. Without loading more specialized modules\footnote{\chg{The set.factory function allows explicit control over sampler factories eg. set.factory("base::Slice", FALSE, "sampler")}} or reparameterizing \cite{papaspiliopoulos2007general} the model to trigger better samplers , this can manifest as very poor mixing that looks like non-convergence.}

  \item \chg{\textbf{RNG and chain-independence issues.}
  Random-number generators and seeds could have an impact on the convergence; chains may share effectively the same random stream or highly correlated streams. This violates the assumption of independent chains and can artificially deflate $\hat{R}$, disguising genuine non-convergence \cite{Gelman2013}. To assess the robustness of posterior inference with respect to random-number generation, particularly in the presence of slow mixing or potential non-convergence, it is recommended to conduct multiple independent runs using different random seeds.\footnote{In \texttt{JAGS}/\texttt{rJAGS}, independent random-number streams can be enforced by specifying chain-specific RNG settings in the initialization function, e.g., \texttt{inits\_fun <- function(chain\_id) \{ list(.RNG.name = "base::Mersenne-Twister", .RNG.seed = 9404 + chain\_id * 16) \}}.}}

\end{itemize}

%% file: app04_fusion.tex
\newpage
\section{Additional reversed model results for illustration 6}
\label{app:dsem_illu6}

\begin{figure}[H]
    \centering
    \includegraphics[width=\textwidth]{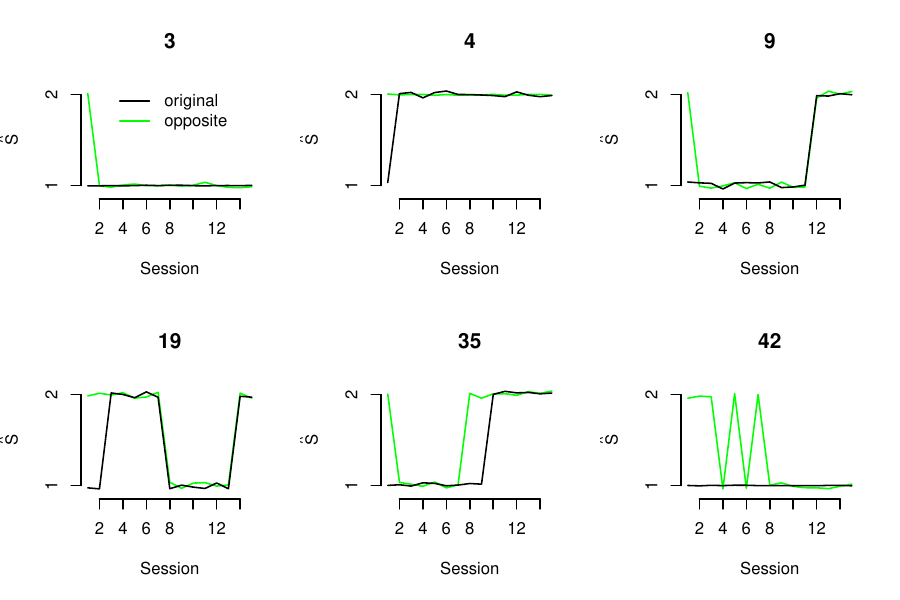}
    \caption{Sample state memberships for 6 patients for the original Fusion model from illustration 6 (black) compared to a model that specifies the opposite mechanism (switch from single factor model to three factor model; green). For the Figure, states in the opposite model were recoded, so that $S_1$ always refers to the three factor state (and vice versa). Note that state-membership was forced to be $S_1$ in illustration 6 and $S_2$ in the opposite model.}
    \label{fig:defus1}
\end{figure}

\begin{figure}
    \centering
    \includegraphics[width=\textwidth]{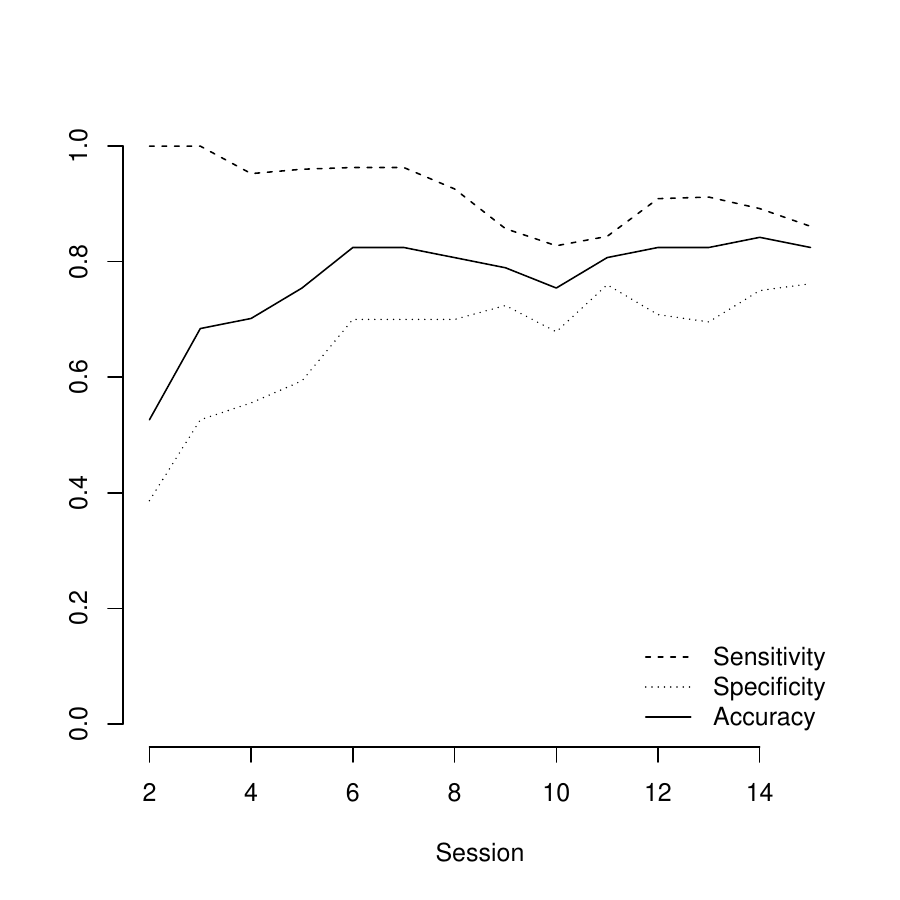}
    \caption{Sensitivity, Specificity, and Accuracy at each session with regard to the recovery of the original Fusion model from illustration 6 compared to a model that specifies the opposite mechanism (switch from single factor model to three factor model). For the Figure, states in the opposite model were recoded, so that $S_1$ always refers to the three factor state (and vice versa).}
    \label{fig:defus2}
\end{figure}